\documentclass[twocolumn]{aastex631}

\usepackage{lineno}

\shorttitle{Insight into the Galactic Bulge Chemodynamical Properties}
\shortauthors{Liao et al.}

\graphicspath{{./}}

\usepackage{graphicx, hyperref}

\usepackage{marvosym}

\usepackage{marvosym}

\begin{document}

\title{Insight into the Galactic Bulge Chemodynamical Properties from Gaia DR3}

\author{Xiaojie Liao}
\affiliation{Department of Astronomy, School of Physics and Astronomy, 800 Dongchuan Road, Shanghai Jiao Tong University, Shanghai 200240, \\China}
 \affiliation{Key Laboratory for Particle Astrophysics and Cosmology (MOE) / Shanghai Key Laboratory for Particle Physics and Cosmology, Shanghai 200240, China}

\author{Zhao-Yu Li\textsuperscript{\Letter}}
\affiliation{Department of Astronomy, School of Physics and Astronomy, 800 Dongchuan Road, Shanghai Jiao Tong University, Shanghai 200240, \\China}
 \affiliation{Key Laboratory for Particle Astrophysics and Cosmology (MOE) / Shanghai Key Laboratory for Particle Physics and Cosmology, Shanghai 200240, China}

\author{Iulia Simion\textsuperscript{\Letter}}
\affiliation{Shanghai Key Laboratory for Astrophysics, Shanghai Normal University, 100 Guilin Road, Shanghai, 200234, China}

\author{Juntai Shen}
\affiliation{Department of Astronomy, School of Physics and Astronomy, 800 Dongchuan Road, Shanghai Jiao Tong University, Shanghai 200240, \\China}
 \affiliation{Key Laboratory for Particle Astrophysics and Cosmology (MOE) / Shanghai Key Laboratory for Particle Physics and Cosmology, Shanghai 200240, China}
 
\author{Robert Grand}
\affiliation{Astrophysics Research Institute, Liverpool John Moores University, 146 Brownlow Hill, Liverpool, L3 5RF, UK}

\author{Francesca Fragkoudi}
\affiliation{Institute for Computational Cosmology, Department of Physics, Durham University, Durham DH1 3LE, UK}

\author{Federico Marinacci}
\affiliation{Department of Physics and Astronomy ”Augusto Righi”, University of Bologna, via Gobetti 93/2, I-40129 Bologna, Italy}
\affiliation{INAF, Astrophysics and Space Science Observatory Bologna, Via P. Gobetti 93/3, I-40129 Bologna, Italy}

\correspondingauthor{Zhao-Yu Li (lizy.astro@sjtu.edu.cn), Iulia Simion (iuliateodorasim@yahoo.com)}

\begin{abstract}
We explore the chemodynamical properties of the Galaxy in the azimuthal velocity $V_\phi$ and metallicity [Fe/H] space using red giant stars from Gaia Data Release 3. The row-normalized $V_\phi$-[Fe/H] maps form a coherent sequence from the bulge to the outer disk, clearly revealing the thin/thick disk and the Splash. The metal-rich stars display bar-like kinematics while the metal-poor stars show dispersion-dominated kinematics. The intermediate-metallicity population ($-1<$[Fe/H]$<-0.4$) can be separated into two populations, one that is bar-like, i.e. dynamically cold ($\sigma_{V_R}\sim80$ $\rm km\ s^{-1}$) and fast rotating ($V_\phi\gtrsim100$ $\rm km\ s^{-1}$), and the Splash, which is dynamically hot ($\sigma_{V_R}\sim110$ $\rm km\ s^{-1}$) and slow rotating ($V_\phi\lesssim100$ $\rm km\ s^{-1}$).  We compare the observations in the bulge region with an Auriga simulation where the last major merger event occurred $\sim10$ Gyr ago: only stars born around the time of the merger reveal a Splash-like feature in the $V_\phi$-[Fe/H] space, suggesting that the Splash is likely merger-induced, predominantly made-up of heated disk stars and the starburst associated with the last major merger. Since the Splash formed from the proto-disk, its lower metallicity limit coincides with that of the thick disk. The bar formed later from the dynamically hot disk with [Fe/H] $>-1$ dex, with the Splash not participating in the bar formation and growth. Moreover, with a set of isolated evolving $N$-body disk simulations, we confirm that a non-rotating classical bulge can be spun up by the bar and
develop cylindrical rotation, consistent with the observation for the metal-poor stars.

\end{abstract}

\keywords{Galaxy: bulge --- Galaxy: Chemistry and kinematics}

\section{Introduction} 
\label{sec:intro}
The physical processes involved in the formation of the Galactic bulge encode important information about the formation and evolution history of the Milky Way (MW) (\citealp[for review]{barbuyChemodynamicalHistoryGalactic2018}). In the early stage of galaxy formation, the bulges form through merging and dissipative collapse \citep{eggenEvidenceMotionsOld1962} and later evolve through secular evolution. Severe dust extinction strongly affects photometric observations of our galaxy's bulge region. However, infrared observations reveal that the MW hosts an elongated triaxial structure with the near-side in the first quadrant \citep{dwekMorphologyNearinfraredLuminosity1995}. Early evidence for the existence of the bulge also exists from gas kinematics \citep{binneyUnderstandingKinematicsGalactic1991} with later observations revealing that the bulge rotates cylindrically \citep{howardKINEMATICSEDGEGALACTIC2009,kunderBULGERADIALVELOCITY2012,zoccaliGIRAFFEInnerBulge2014}, a signature of a rotating rigid body. 
Our current view is that the MW hosts a boxy/peanut-shaped bulge or pseudo-bulge \citep{maihara4MicronObservationGalaxy1978,weilandCOBEDiffuseBackground1994,dwekMorphologyNearinfraredLuminosity1995,lopez-corredoiraBoxyBulgeMilky2005}, consistent with a buckled bar formed from the secular evolution of a disk. The presence of a small classical bulge in the MW, i.e., a merger-induced spherical structure similar in many aspects to a mini-elliptical galaxy, has not been excluded but its mass is likely less than 8\% of the disk mass \citep{shenOURMILKYWAY2010a}.

The Galactic bulge exhibits complex chemodynamical trends that have been revealed by spectroscopic surveys such as the  BRAVA \citep{richBulgeRadialVelocity2007, kunderBULGERADIALVELOCITY2012}, ARGOS \citep{freemanARGOSIIGalactic2013}, GIBS \citep{zoccaliGIRAFFEInnerBulge2014}, GES \citep{rojas-arriagadaGaiaESOSurveyMetallicity2014,rojas-arriagadaGaiaESOSurveyExploring2017}, APOGEE \citep{majewskiApachePointObservatory2017,jonssonAPOGEEDataSpectral2020}, the PIGS \citep{arentsenPristineInnerGalaxy2020} and the BDBS survey \citep{limBlancoDECamBulge2021}. The metallicity distribution function in the bulge region spans a wide range of metallicities and appears to have multiple peaks \citep{richSpectroscopyAbundances881988,mcwilliamFirstDetailedAbundance1994,hillMetallicityDistributionBulge2011,fulbrightAbundancesBaadeWindow2006,zoccaliMetalContentBulge2008a,nessARGOSIIIStellar2013,gonzalezGIRAFFEInnerBulge2015,nessMetallicityDistributionMilky2016,zoccaliGIRAFFEInnerBulge2017,rojas-arriagadaGaiaESOSurveyMetallicity2014,rojas-arriagadaGaiaESOSurveyExploring2017,rojas-arriagadaHowManyComponents2020}, indicating the existence of different stellar populations in the bulge. Several works have confirmed a negative vertical metallicity gradient in the bulge \citep{minnitiMetallicityGradientGalactic1995,zoccaliMetalContentBulge2008a,nessARGOSIVKinematics2013,haydenChemicalCartographyAPOGEE2015}. \cite{nessARGOSIVKinematics2013} attributes this vertical metallicity gradient to the varying fraction of different stellar populations, which have almost fixed mean metallicity and metallicity dispersion, with distance from the galactic plane.

Galactic archaeology studies the kinematical and chemical properties of stars in order to reconstruct the history of formation of our Galaxy since the cosmic dawn. In the Galactic bulge region, chemo-kinematic studies have revealed that the different stellar populations, separated by their metallicity, have different kinematics \citep{nessARGOSIVKinematics2013,clarksonChemicallyDissectedRotation2018,arentsenPristineInnerGalaxy2020,wylieA2A210002021}. Stars with [Fe/H] $>-1$ dex are believed to be originated from disk material as they exhibit comparable cylindrical rotation and low velocity dispersion \citep{kunderBULGERADIALVELOCITY2012,nessARGOSIVKinematics2013, dimatteoDiscOriginMilky2016a, fragkoudiDiscOriginMilky2018} and there seems to be a chemical similarity between these bulge stars and thick disk stars \citep{rojas-arriagadaGaiaESOSurveyExploring2017}.
Metal-poor stars  with [Fe/H] $\lesssim-1$ dex, which exhibit high velocity dispersion \citep{arentsenPristineInnerGalaxy2020,dekanyVVVSurveyNearinfrared2013}, have an unclear origin: they may be either formed in situ, e.g. ancient in-situ stars \citep{10.1007/10719504_62}, halo interlopers \citep{luceyCOMBSSurveyII2020}, Aurora (\citealp[i.e., ancient in-situ stars]{belokurovDawnTillDisc2022}) or ex-situ e.g., remnants of early accretion events \citep{hortaEvidenceAPOGEEPresence2020}. More recently, \cite{marchettiBlancoDECamBulge2023} used 2.3 million red clump stars in the Blanco DECam Bulge survey (BDBS) to study the chemo-kinematic properties of the Galactic bulge. They found that metal-rich stars ([Fe/H] $>-0.5$ dex) show bar signatures, with the metal-poor stars exhibiting isotropic motions.

 The classic picture of bulge formation requires a dynamically cold disk in which a bar instability can be triggered. Recent numerical simulations were able to demonstrate that a bulge can form also from a hot thick disk \citep{ghoshBarsBoxyPeanut2023}. Disks at high redshifts ($z\sim4-5$) have been observed by ALMA \citep{roman-oliveiraRegularRotationLow2023,parlantiALMAHintsPresence2023} and JWST \citep{ferreiraJWSTHubbleSequence2023} and even bars have been clearly observed at $z>2$ \citep{guoFirstLookBars2023,leconteJWSTInvestigationBar2023,costantinMilkyWaylikeBarred2023}. At early times, the emergence of stellar bars can be rapid when dark matter fraction in the centres of disk galaxies is low \citep{bland-hawthornRapidOnsetStellar2023}. In this work we will further show, using high redshift snapshots of a Milky-Way like simulation, that it is possible to form a bar immediately after the last major merger $\sim10$ Gyr ago when the disk was still dynamically hot.

The European Space Agency's Gaia satellite has revolutionized our understanding of the MW's past, with the second and third data releases (DR3) revealing plenty of previously unknown galactic features \citep{gaiacollaborationGaiaDataRelease2018,gaiacollaborationGaiaEarlyData2021,antojaDynamicallyYoungPerturbed2018}. Gaia's astrometric measurements combined with radial velocities from the Radial Velocity Spectrometer (RVS) or ground-based spectroscopic surveys enabled the study of individual stars in the orbital--chemical space, towards the Galactic bulge \citep{queirozMilkyWayBar2021}. Using Gaia and APOGEE chemical abundance measurements, \cite{belokurovDawnTillDisc2022} conducted a deep analysis of the Galactic disk in the $V_{\phi}-{\rm [Fe/H]}$ space. In the aluminium-selected `in-situ' population, they identified an ancient metal-poor population that is kinematically hot with an isotropic velocity ellipsoid and modest net rotation (dubbed `Aurora'), and a subsequent `spin-up' phase of the disk stars which culminates with the formation of the Splash. The Splash \citep{belokurovBiggestSplash2020} is a relatively metal-rich population ([Fe/H] $>$ -0.7 dex) with low azimuthal velocity dispersion (50-60 $\rm km\ s^{-1}$), likely originating from the proto-disk, heated by the Gaia Sausage Enceladus \citep[GSE,][]{helmiMergerThatLed2018, 2018MNRAS.478..611B} merger event. The presence of a kinematically heated disc component in the halo as an imprint left by an accretion event had already been suggested by \cite{dimatteoMilkyWayHas2019}. 

In this work, we aim to look for evidence of the Splash and the bar/bulge in the Galactic bulge region, in an attempt to decipher the origin of the chemodynamical properties of the Galactic bulge.
For this goal, we use a recently released catalog based on Gaia-XP spectra that reaches the bulge \citep{andraeRobustDatadrivenMetallicities2023}. We describe the catalog in Section~\ref{sec:data} and analyze the observations in Section~\ref{sec:chemores}, with a focus on the $V_{\phi}-$[Fe/H] space. In Section~\ref{sec:cossim},\ref{sec:nbody}, we employ the Auriga \citep{grandAurigaProjectProperties2017} cosmological simulation and three $N-$body simulations to better understand the chemodynamical properties of the bulge region as revealed by our sample. In Section~\ref{sec:dis} we discuss our results and in Section~\ref{sec:con} we summarize our findings. 

\begin{figure}[ht!]
	\includegraphics[width=78mm]{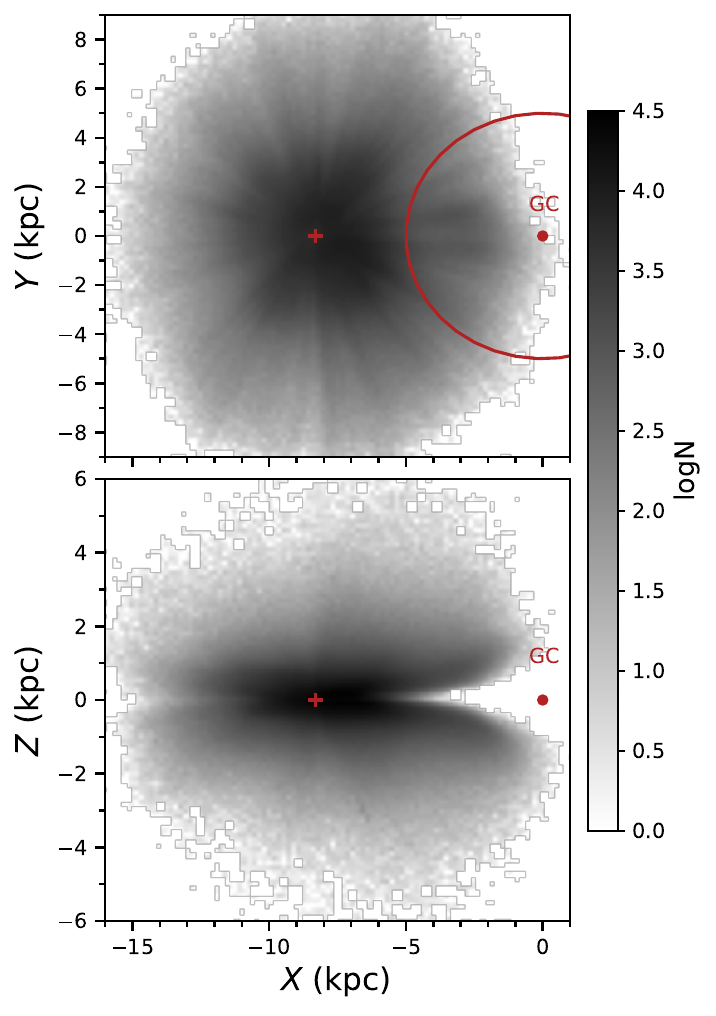}
	\caption{Spatial distribution of the full Gaia-XP-spectra-derived sample (5,497,737 stars, sample A), projected on the $X-Y$ (top) and $X-Z$ (bottom) planes. The red cross and dot denote the Sun position and the Galactic center (GC), respectively. The red circle of 5 kpc radius encloses the bulge stars (330,414, sample B). \label{fig:fig1}}
\end{figure}

\section{Data} \label{sec:data}

\begin{figure}[ht!]
	\includegraphics[width=87mm]{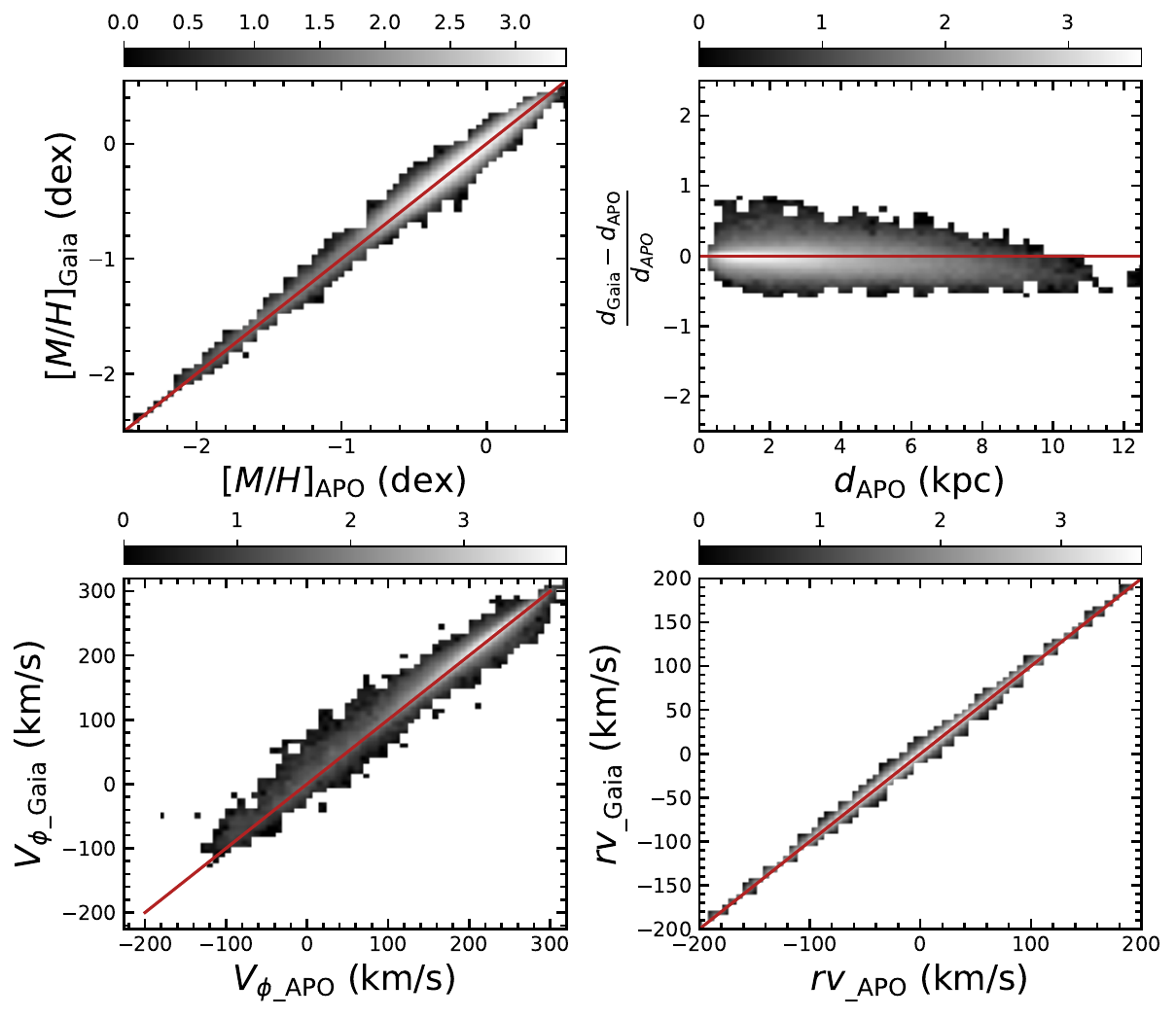}
	\caption{Comparison of the metallicity, distance, azimuthal velocity and radial velocity for the common stars in the APOGEE DR17 and Gaia samples. The two samples are in good agreement. The distances adopted for Gaia and APOGEE are Bayesian distance \citep{bailer-jonesEstimatingDistancesParallaxes2021} and StarHorse distance \citep{queirozBulgeOuterDisc2020}, respectively. \label{fig:fig2}}
\end{figure}

In this work, we use a catalog of 13.3 million red giants with stellar metallicity from \cite{andraeRobustDatadrivenMetallicities2023}. The metallicity is derived from the low-resolution XP spectra in Gaia DR3 \citep{gaiacollaborationGaiaDataRelease2022} using the data-driven method XGBoost, that is trained on 510,413 APOGEE stars \citep{majewskiApachePointObservatory2017} augmented by 291 ultra-metal-poor stars from LAMOST \citep{liFourhundredVeryMetalpoor2022}. According to \cite{andraeRobustDatadrivenMetallicities2023}, the typical uncertainty on metallicity is 0.1 dex. We select stars with radial velocity errors smaller than 2.0 $\rm km\ s^{-1}$ and \verb| parallax/parallax_error > 5|, which results in a sample of 5,497,737 stars (sample A). We use the photogeometric distances calculated by \cite{bailer-jonesEstimatingDistancesParallaxes2021} to complete the 6D phase-space information. The six-dimensional Gaia observables are transformed to Galactocentric cylindrical coordinates using Astropy\footnote{\url{https://www.astropy.org/index.html}} \citep{robitailleAstropyCommunityPython2013, theastropycollaborationAstropyProjectBuilding2018, theastropycollaborationAstropyProjectSustaining2022} with the Sun position at ($X$, $Y$, $Z$) $=$ (-8.34, 0, 0.027) kpc \citep{reidTRIGONOMETRICPARALLAXESHIGH2014} and peculiar motion ($V_X$, $V_Y$, $V_Z$) $=$ (11.1, 12.24, 7.25) $\rm km\ s^{-1}$ \citep{schonrichGalacticRotationSolar2012a} in a right-handed frame. The spatial distribution of this sample is shown in Fig.~\ref{fig:fig1}. From the figure it is obvious that the Gaia data only covers the near side of the bulge (distance to the Sun less than 8.3 kpc) and almost avoids the low latitude region. We do see an overdensity in the bulge region within the red circle, confirming the coverage of the bulge stars in our sample.

\begin{figure*}[ht!]
	\includegraphics[width=179mm]{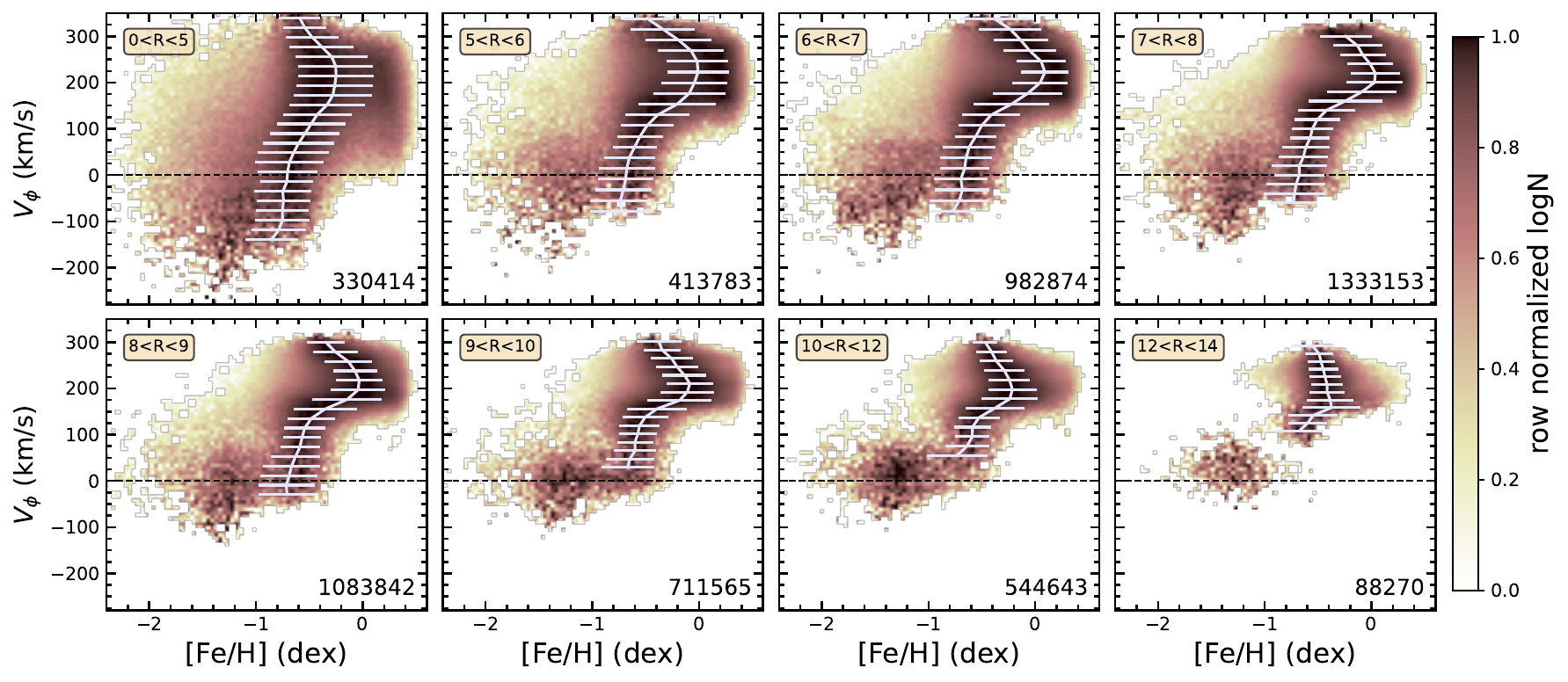}
	\caption{Row-normalized number density maps of the $V_\phi$-[Fe/H] space of stars at various radial bins indicated at the top left corner of the panel. The white curve in each panel represents the best-fit mean [Fe/H] values by a Gaussian-Hermite function at a sequence of small $V_\phi$ bins (see texts for detail), with the horizontal white lines indicating the corresponding 1$\sigma$ metallicity dispersion within each $V_\phi$ bin (stars with [Fe/H]$<-1$ dex are not included). The number at the bottom right corner of each panel denotes the number of stars in that radial bin. The horizontal dashed black line indicates $V_\phi=0$. \label{fig:fig3}}
\end{figure*}

For consistency checks, in Fig.~\ref{fig:fig2} we compare the metallicity, azimuthal and radial velocities of stars common between the Gaia and APOGEE DR17 surveys. We also compare the APOGEE-based StarHorse distances \citep{queirozBulgeOuterDisc2020} with Gaia-based Bayesian distances \citep{bailer-jonesEstimatingDistancesParallaxes2021} in the upper right panel of the figure. The radial velocities, azimuthal velocities and metallicities are in good agreement while the Gaia distances are slightly under-estimated compared to the StarHorse distances. As mentioned earlier, \cite{andraeRobustDatadrivenMetallicities2023} primarily trained the Gaia XP spectra on APOGEE, therefore we expect good agreement between the two surveys for the common stars. The cylindrical azimuthal velocities are computed using Gaia proper motions for both surveys, but Gaia and StarHorse-derived distances respectively. The bulge sample we analyze in the following sections is selected to be within 5 kpc of Galactocentric radius from the GC (see red circle in Fig.~\ref{fig:fig1}), decreasing the sample A size to 330,414 stars (sample B).

\section{Results} \label{sec:chemores}

\subsection{$V_\phi$-[Fe/H] Maps from Disk to Bulge Region} \label{subsec:dtob}

The morphological, kinematical and chemical properties of the bulge are mainly shaped via its formation and evolution processes; the merger-induced classical bulge is spheroidal-like and dispersion-dominated, while the disk-instability-induced pseudo-bulge is disk-like and rotation-dominated \citep{kormendySecularEvolutionFormation2004, kormendySecularEvolutionDisk2008}. \cite{shenOURMILKYWAY2010a} demonstrated that the Galactic bulge is purely disk-originated via a simple but realistic $N$-body model which reproduces the stellar kinematics of the BRAVA survey \citep{richBulgeRadialVelocity2007} and the photometric asymmetric boxy shape of the Galactic bulge \citep{dwekMorphologyNearinfraredLuminosity1995, weilandCOBEDiffuseBackground1994,lopez-corredoiraBoxyBulgeMilky2005}  remarkably well. Therefore, the chemodynamical properties of many bulge stars should closely resemble those of disk stars. 

\begin{figure}[ht!]
	\includegraphics[width=84mm]{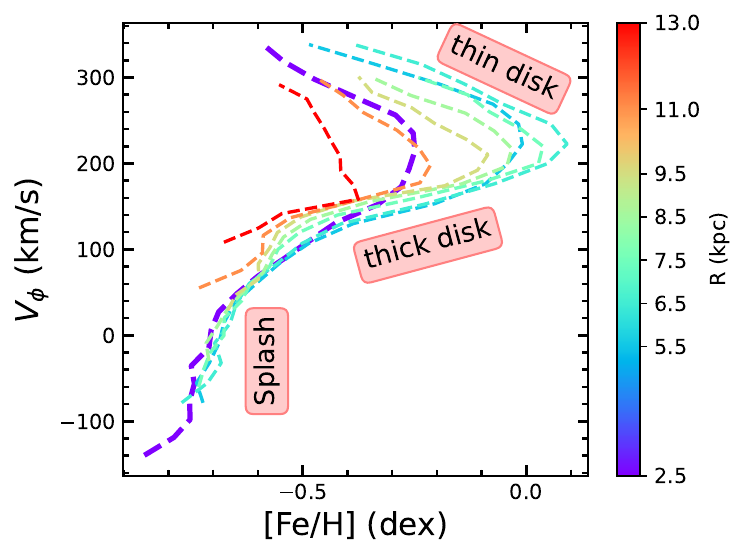}
	\caption{$V_\phi$-[Fe/H] patterns at different radii from Figure~\ref{fig:fig1} (white curves) are shown here together to better illustrate their systematic variation across the Galactic disk. The color indicates the radial bin, with the bulge region (R $<$ 5 kpc) shown in purple. \label{fig:fig4}}
\end{figure}

To explore the possible chemodynamical connection between the bulge and disk stars, we follow the approach of \cite{belokurovBiggestSplash2020} and build row-normalized number density maps of our sample A (see Section \ref{sec:data}) in the $V_\phi$-[Fe/H] space for consecutive radial annuli, as shown in Fig.~\ref{fig:fig3}. In order to better quantify the general trend of the $V_\phi$-[Fe/H] distribution across the Galactic disk, for each radial bin we split the subsample into small $V_\phi$ bins and fit a Gaussian-Hermite function to their metallicity distribution profile in order to obtain a sequence of best-fit mean [Fe/H] values and dispersions, which we overlay in the white curve in Fig.~\ref{fig:fig3}. In this procedure we do not include stars with [Fe/H] $<-1$ dex that are believed to be mostly accreted stars, halo stars or in-situ Aurora stars. We observe a similar trend in all panels, from the outer disk to the inner few kpc: a chevron-like pattern at positive azimuthal velocities with a vertical extension towards small and negative azimuthal velocities. The upper and lower branches of the chevron-pattern  and the vertical extension, have been suggested to trace the thin disc, the thick disk and the Splash, respectively \citep{belokurovBiggestSplash2020,belokurovDawnTillDisc2022,leeChemodynamicalAnalysisMetalrich2023}. The upper branch of the chevron, with high rotational velocities is overall more metal-rich, while the lower branch with intermediate rotational velocity is more metal-poor, in agreement with our knowledge of the thin and thick disks. As for the vertical extension towards lower $v_{\phi}$, it is consistent with the Splash, i.e., a heated disk population or the in-situ halo. Note that the density map in Fig.~\ref{fig:fig3} is row-normalized: in the Splash region, the number of prograde stars actually outweighs that of retrograde stars, resulting in an apparent small net rotation. 



Although the chevron pattern looks similar at different radii, there is a systematic variation across the disk in the slopes and conjunction points of the thin and thick disk branches. To better illustrate it, we draw the sequences (the white curves in Fig.~\ref{fig:fig3}) together in Fig.~\ref{fig:fig4}, where each color now indicates the radial bin. Firstly, we notice that the vertical extensions (corresponding to the Splash) at different radii overlap and all have similar peak metallicity, indicating that they have a common origin in a global event, i.e., the GSE merger. Secondly, the chevron patterns form a coherent sequence with the peak [Fe/H] decreasing with radius, except for the innermost radial bin ($R<5$ kpc, see purple line Fig.~\ref{fig:fig4}). By extrapolating the trends revealed by the profiles of the outer radial bins, we would expect the chevron-pattern of the innermost radial bin (sample B) to be located at the highest metallicities. However, it is located in the middle of the other profiles, with a conjunction point at [Fe/H] $\approx$ -0.25 dex. This could be partially caused by the obeservational bias: there are almost no stars in our sample close to the Galactic plane due survey-specific limitations, in particular the strong dust extinction. Since stars closer to the mid-plane are more metal-rich, the absence of disk stars would shift the overall metallicity trend towards the more metal-poor region. Moreover, those relatively metal-poor stars with metallicities larger than -1 dex of different origin (accreted stars, halo interlopers, metal-poor pre-existing in- situ stars, etc.) are also present, further shifting the overall metallicity trend to lower [Fe/H] values.

For the thin disk population, the rotational velocity $V_\phi$ is anti-correlated with metallicity. In a scenario in which the MW forms inside-out \citep[e.g.][]{frankelInsideoutGrowthGalactic2019}, this trend is mainly due to radial migration through churning and blurring \citep{schonrichUnderstandingInverseMetallicity2017a, vickersFlatteningMetallicityGradient2021}, and the negative metallicity gradient of the thin disk stars: at a specific radius, stars with smaller $V_\phi$ likely migrated from the inner region that is more likely metal-rich, giving rise to the negative correlation between $V_\phi$ and [Fe/H]. On the other hand, for the thick disk population, $V_\phi$ is positively correlated with [Fe/H]. Thick disk stars are older and as such they had more time to experience perturbations, resulting in larger velocity dispersion and slower rotation; because of their less circular orbits and higher distances from the Galactic plane $z$, they also experienced less radial migration \citep{vera-ciroEffectRadialMigration2014}.

From Fig.~\ref{fig:fig4} it is also apparent that the chevron-patterns shift towards lower metallicities as the radius increases. The intervals between the upper branches (the thin disk) of the chevron-like pattern at different radii are  larger than the intervals of the lower branches (the thick disk), which are almost aligned with each other; the intervals reflect the steeper negative metallicity radial gradient of the thin disk, and the weaker metallicity gradient in the thick disk, seemingly confirming that the thick disk is more disturbed and well-mixed than the thin disk, with a weaker metallicity gradient \citep{vickersFlatteningMetallicityGradient2021}. The negative metallicity gradient of the thin disk can be reproduced in an inside-out disk formation scenario in cosmological simulations \citep{valentiniChemicalEvolutionDisc2019}.
In addition, the slopes of the upper branches (thin disk) increase with radius $R$. This may also be explained with an inside-out formation scenario, where the inner disk has more time to enrich itself, thus becoming more metal-rich and with a broader metallicity distribution than the outer regions. Considering the flat rotation curve and the asymmetric drift effect, the slope of the $V_{\phi}-$[Fe/H] profile of the outer disk is expected to be steeper than the inner disk. The slopes of the lower branches (thick disk) are shallower and almost constant with radius, implying that the thick disk could have formed in one single episode within a short time-frame, an event which likely occurred in the turbulent early epoch when the gas supply from mergers was abundant and well mixed.
\begin{figure}[ht!]
    \centering
	\includegraphics[width=80mm]{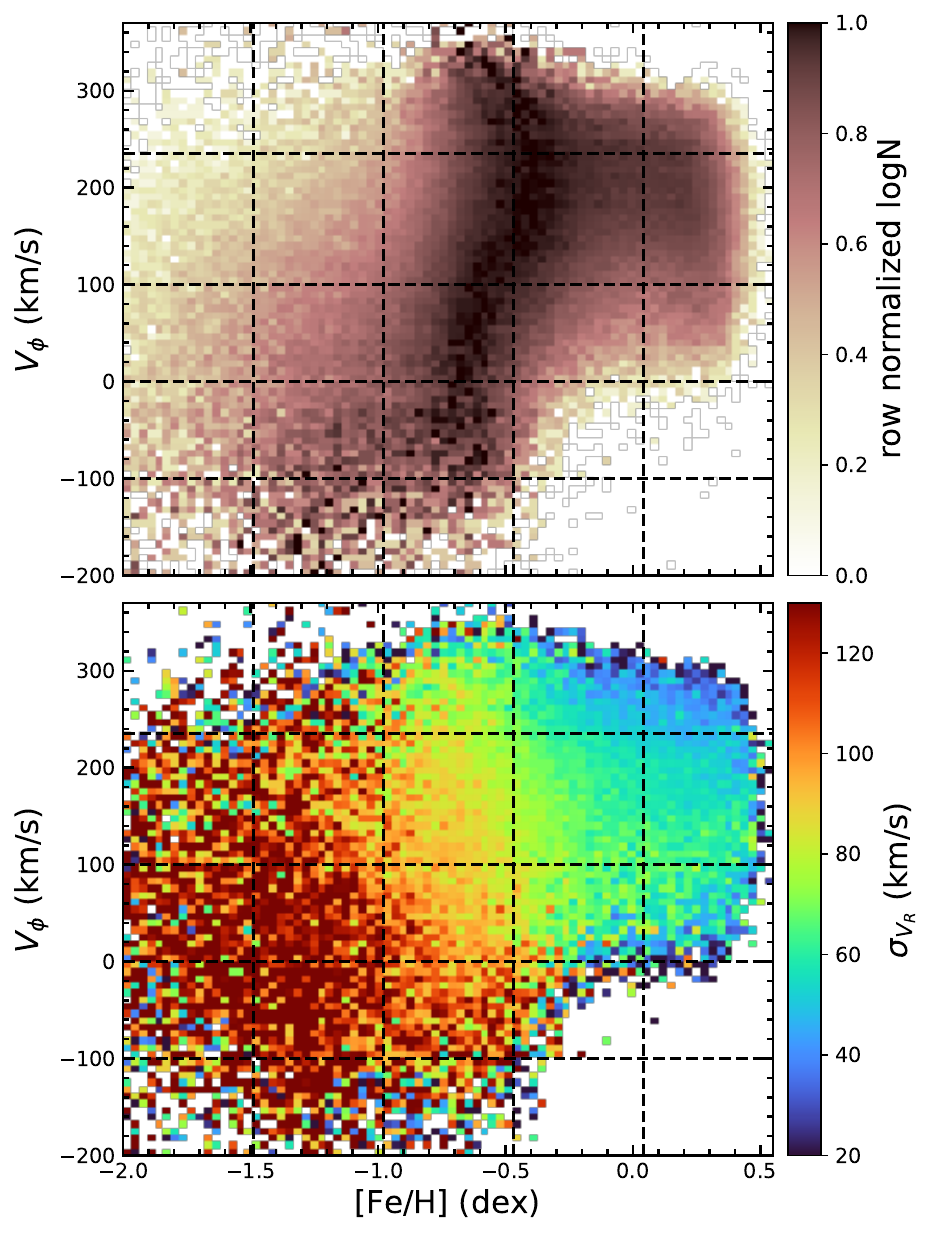}
	\caption{Top panel: The row normalized number density of bulge stars ($R<5$ kpc, sample B) in $V_\phi$-[Fe/H] space with the dashed lines indicating the division of stars into different grids. Bottom panel: similar to the top panel but color-coded with the radial velocity dispersion. A clear pattern emerges with higher dispersion at the bottom left corner and lower dispersion at the top right corner. \label{fig:fig5}}
\end{figure}

\subsection{Chemodynamical Properties of the Galactic Bulge} \label{subsec:pops}
In the last subsection, similar general trends in the $V_\phi$-[Fe/H] space have been revealed for both the bulge and disk stars. In this section, our focus turns to the bulge region (sample B, introduced in Section 2).

\begin{figure}[ht!]
    \centering
	\includegraphics[width=85mm]{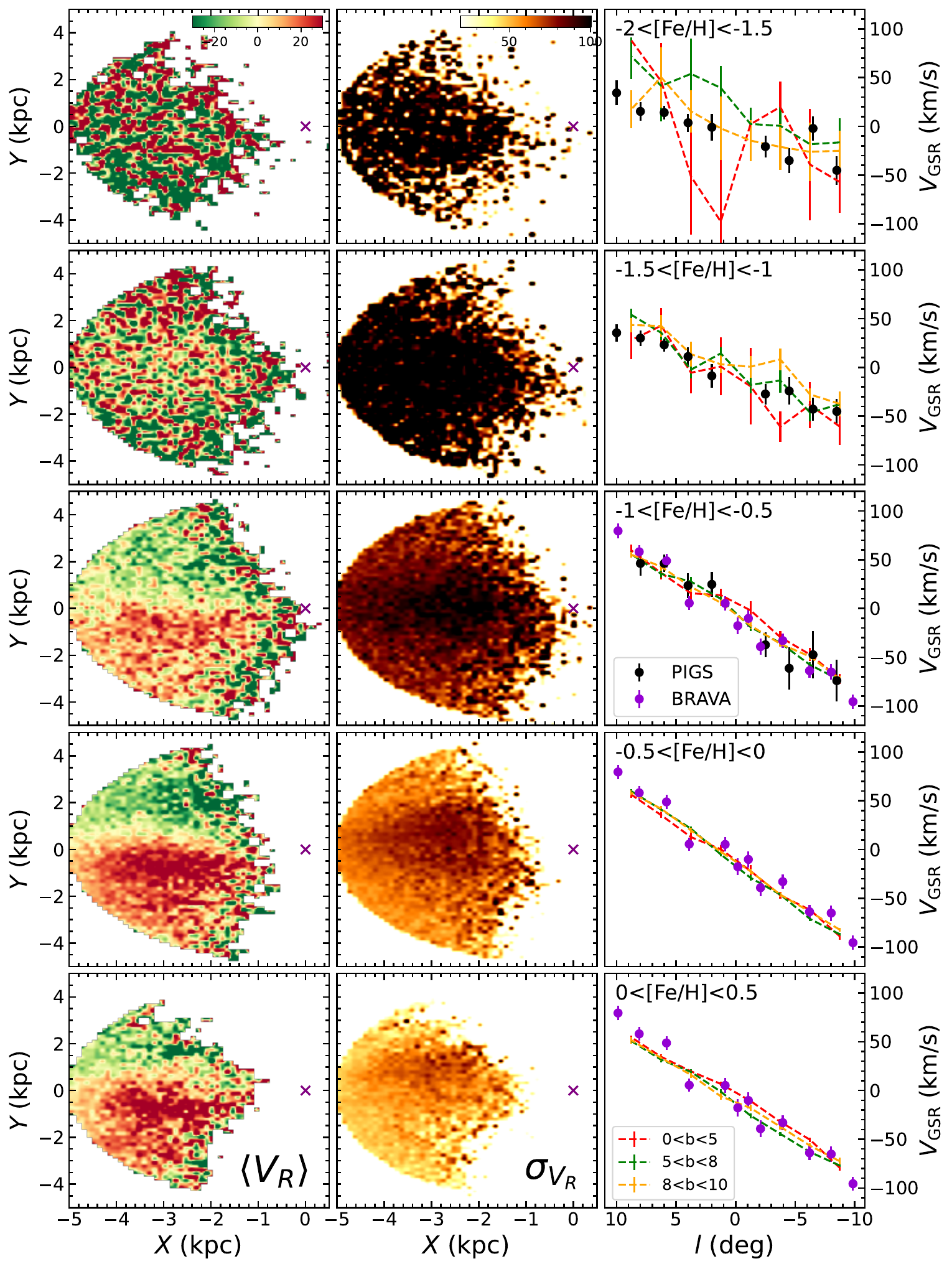}
	\caption{Kinematic properties of the Galactic bulge stars (sample B) within different metallicity bins sliced as the dashed vertical lines marked in Fig.~5. The mean radial velocity $\left\langle V_R \right\rangle$ (left column) and  radial velocity dispersion $\sigma_{V_R}$ (middle column) are projected on the $X-Y$ plane. The position of the GC is marked with a purple cross. The rotation profiles ($l-V_{gsr}$) at different latitudes of 0$^\circ$ $<$b$<$ 5$^\circ$ (red), 5$^\circ <$ b $<8^\circ$ (green) and 8$^\circ<$ b $<$10$^\circ$ (yellow) are shown in the right column. The PIGS survey \citep[black points,][]{arentsenPristineInnerGalaxy2020} and the BRAVA survey \citep[blue points,][]{kunderBULGERADIALVELOCITY2012} show good agreement with the Gaia sample.\label{fig:fig6}}
\end{figure}

In the disk region (R $>$ 5 kpc), the $V_\phi$-[Fe/H] pattern can be attributed to the thin, thick disks and the Splash. However, in the bulge region (R $<$ 5 kpc), it is more complicated to attribute this pattern to distinct stellar populations, e.g. the bar, the Splash, Aurora, or the inner disks. The bar formation is a key event in the history of the MW, which requires a pre-existing stellar disk. Evidence of the bar in the $V_\phi$-[Fe/H] space could help revealing the possible formation epoch of the Galactic disk.  

We show the row normalized number density map in the $V_\phi$-[Fe/H] space for the bulge sample in the top panel of Fig.~\ref{fig:fig5} (note that it is just the enlarged version of the top-left panel in Fig.~\ref{fig:fig3}) and the radial velocity dispersion $\sigma_{V_R}$ in the bottom panel. The radial velocity dispersion increases from the upper right corner to the lower left corner monotonically. 

\begin{figure*}[ht!]
    \centering
	\includegraphics[width=158mm]{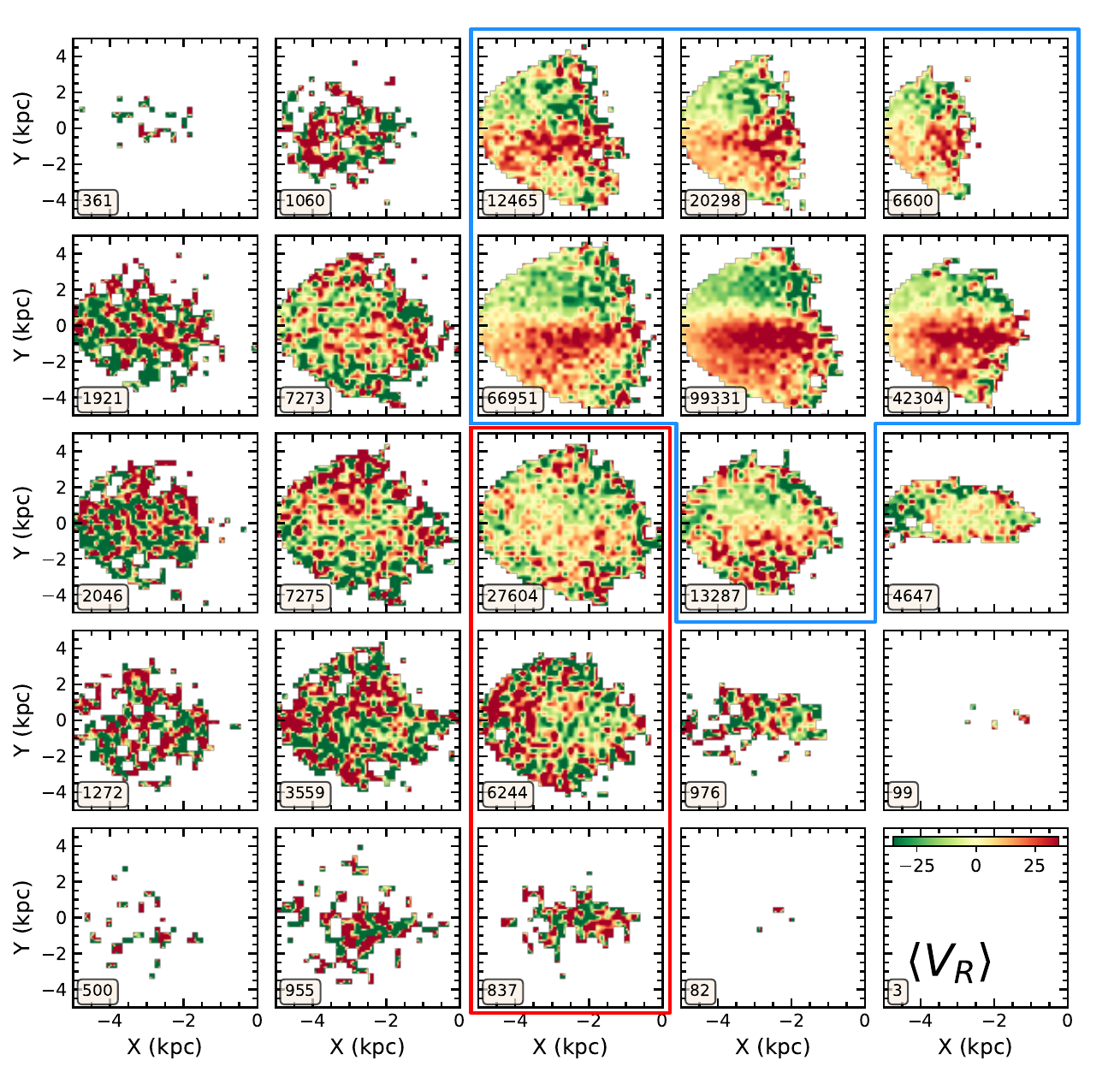}
	\caption{The $\left\langle V_R \right\rangle$ maps in $X-Y$ plane for the grids defined in the $V_\phi$-[Fe/H] space in Fig~\ref{fig:fig5}. The position of the panel in this figure corresponds to the position of the grid in Fig.~\ref{fig:fig3}. The number in the bottom left corner of each panel is the number of stars. The panels with good number statistics showing bar-like kinematics are enclosed within the blue frame, consistent with the dynamically cold region in the $\sigma_{V_R}$ map in the bottom panel of Fig~\ref{fig:fig3}. The panels denoting the Splash region are delimited by the red box. No bar-like kinematics can be seen in the Splash and the other more metal-poor stars. \label{fig:fig7}}
\end{figure*}

In previous works, the metallicity is usually the only criterion used to separate different stellar populations in the Galactic bulge region \citep{nessARGOSIIIStellar2013, nessARGOSIVKinematics2013, rojas-arriagadaHowManyComponents2020, wylieA2A210002021}. However, in the Splash region ($-1<$ [Fe/H] $<-0.5$), the dynamically cold ($V_\phi > 100$) and hot ($V_\phi < 100$) populations co-exist in the same metallicity range: the traditional metallicity cut is not accurate enough for dissecting the different structures - kinematics plays a crucial role. Nonetheless, for comparison with previous works, we first examine the kinematic properties of populations with different metallicities. The results are shown in Fig.~\ref{fig:fig6} with the metallicity specified in the upper left corner of the third column. The left two columns show the average radial velocity $\left\langle V_R \right\rangle$ and the radial velocity dispersion $\sigma_{V_R}$ maps in the $X-Y$ plane, respectively. The more metal-rich populations, with [Fe/H] $>-1$ dex, all exhibit `butterfly'-like patterns in the average radial velocity maps, a signature of bar-like kinematics \citep{bovyLifeFastLane2019b,fragkoudiChemodynamicsBarredGalaxies2020a,queirozMilkyWayBar2021,drimmelGaiaDataRelease2023}. They are caused by the positive and negative radial velocities of stars on the bar-supporting orbits in adjacent quadrants \citep{zoccaliGIRAFFEInnerBulge2014, bovyLifeFastLane2019b, drimmelGaiaDataRelease2023}. The corresponding $\sigma_{V_R}$ maps show larger amplitudes closer to the bar major axis (zero average radial velocity) where the radial velocity $V_R$ changes its sign. On the other hand, the metal-poor populations with [Fe/H] $<-1$ dex do not show bar-like kinematics in the $\left\langle V_R \right\rangle$ maps, and their $\sigma_{V_R}$ dispersion maps are quite isotropic, suggesting that they are not part of the bar structure but representing a dispersion-supported structure. Notice that the zero radial velocity line in the $\left\langle V_R \right\rangle$ map should align with the major axis of the bar rather than the $X$-axis (i.e., the Sun-GC line). This is mainly caused by the distance errors, which will be discussed in Section~\ref{sec:dis}. 

The rotation profiles of the bulge sample at various latitude slices are shown in the right column of Fig.~\ref{fig:fig6}, where the errors were obtained via bootstrapping. We overlay data from the PIGS survey \citep[black points,][]{arentsenPristineInnerGalaxy2020} and the BRAVA survey \citep[blue points,][]{shenOURMILKYWAY2010a} at the corresponding metallicity. There is good agreement between Gaia and the previous surveys, highlighting the quality of the Gaia DR3 data even in the bulge region. Except the most metal poor population, all stars in sample B display cylindrical rotation with the more metal-rich stars, [Fe/H] $>-1$ dex, rotating faster than the metal-poor stars, with [Fe/H] $<-1$ dex. It is interesting that the metal-poor stars ($-1.5<$ [Fe/H] $<-1$) display cylindrical rotation, albeit slower than the metal-rich ones and without `butterfly'-like pattern: it has been suggested that a pre-existing low-mass classical bulge spun up by the rotating bar can also develop cylindrical rotation, evidence of secular evolution driven by the bar \citep{sahaSpinupLowmassClassical2012}. This effect will be investigated into more detail in Section \ref{sec:nbody}. 

Recent works have shown that the spin-up phase of the MW disk ends at [Fe/H] $\sim -1$ dex and the bulk of the disk stars form afterwards, with the azimuthal velocity continuously increasing towards higher metallicity \citep{belokurovDawnTillDisc2022, semenovFormationGalacticDisks2023}. Fig.~\ref{fig:fig6} shows that stars in the Galactic bulge region with [Fe/H] $\gtrsim -1$ dex exhibit bar-like kinematics, implying that the primordial inner disk at the time when the bar formed should have similar metallicity property (i.e., [Fe/H] $\gtrsim -1$ dex). The result in the Galactic bulge is consistent with previous works focusing on the Galactic disk \citep{belokurovDawnTillDisc2022}.

To better understand the chemodynamical properties of the Galactic bulge and to disentangle the different stellar populations, we divide the $V_\phi$-[Fe/H] space into a grid (dashed black lines in Fig.~\ref{fig:fig5}) of small cells. In this way, we do not assign the different loci of the $V_\phi$-[Fe/H] diagram to different stellar populations, but investigate systematically each cell in order to avoid a priori bias; in Fig.~\ref{fig:fig7} we show the mean radial velocity maps in the $X-Y$ plane for the stars in each grid cell. Neglecting the panels with low number statistics (the number of stars in each cell is specified in the bottom-left corner), we can see that only some of the remaining panels show bar-like kinematics (\textbf{enclosed within the blue frame}). This result is consistent with the radial velocity dispersion map (bottom panel of Fig.~\ref{fig:fig5}), where the stars with bar-like kinematics have lower radial velocity dispersion ($\sigma_{V_R} \lesssim 90$ $\rm km\ s^{-1}$), i.e., dynamically cold. The panels in the middle column all have the same metallicity range of $[-1, -0.4]$ dex: the top two panels at higher $V_\phi$ and lower $\sigma_{V_R}$ display bar-like kinematics while the bottom three panels corresponding to the Splash region (enclosed with the red line) display dispersion-dominated kinematics. At even lower metallicities (the left two columns of Fig.~\ref{fig:fig7} with $\mathrm{[Fe/H]}\lesssim-1$ dex), the kinematics is consistent with a dispersion-dominated population (i.e., accreted stars, classical bulge, halo stars, or Aurora). The kinematics is therefore crucial to disentangle the bar from other populations (e.g. the Splash) in the bulge region.

\begin{figure}[ht!]
	\includegraphics[width=84mm]{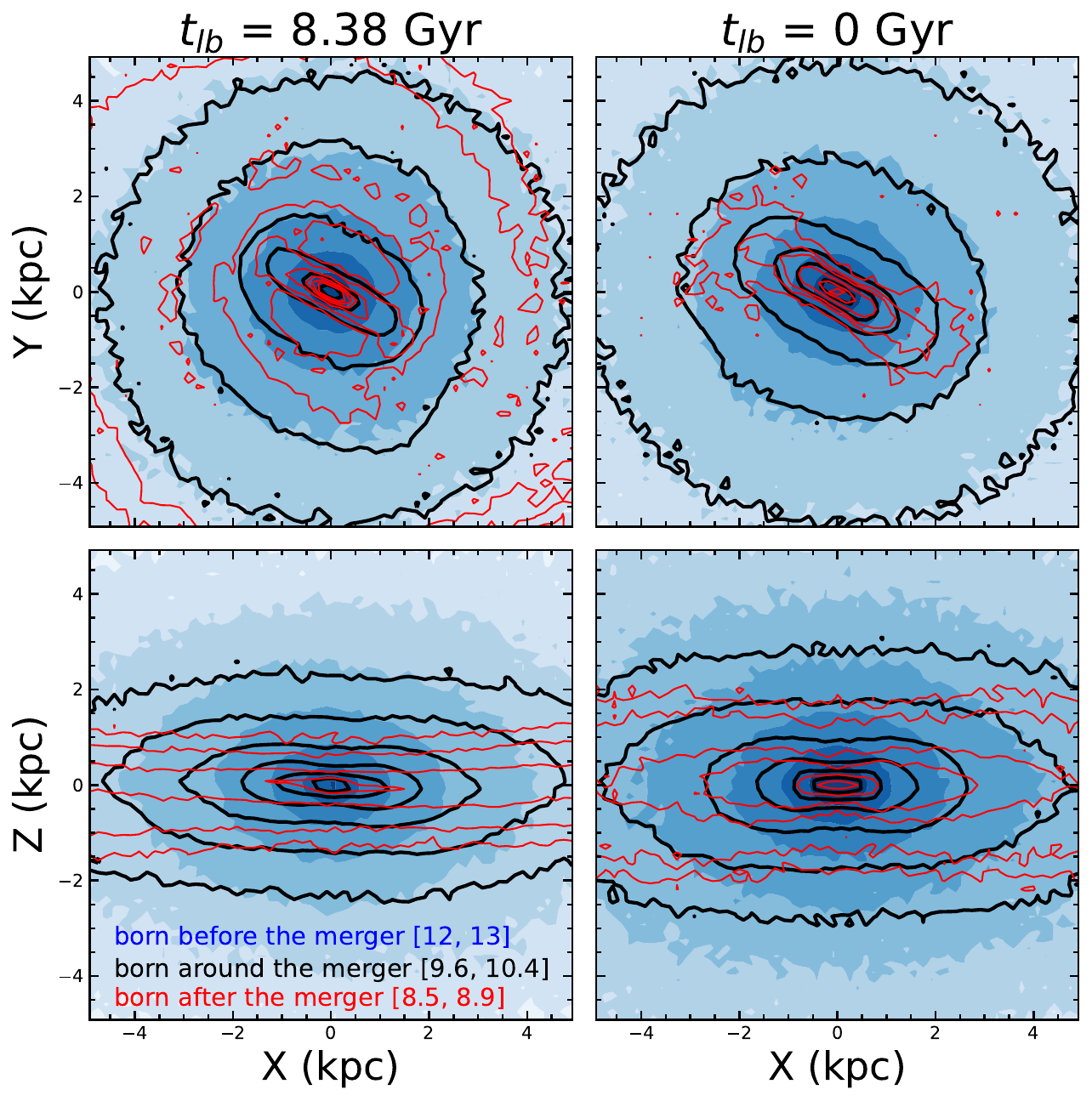}
	\caption{Evolution of the spatial distribution of stars born before, during, and after the last major merger about 10 Gyr ago in the Auriga simulation Au23. The top and bottom rows show the number density maps in the face-on ($X-Y$ plane) and edge-on ($X-Z$ plane) views, respectively. The number density contours of the three populations are shown in different colors, which are indicated in the lower left panel. The left column shows the snapshot with look back time $t_{lb}=$ 8.38 Gyr, which is $\sim1.5$ Gyr after the last major merger. The snapshot at present ($t_{lb}=0$) is shown in the right column. The stars born  before the merger maintain a dynamically hot spherical structure, while the stars born around and after the merger resemble the thick and thin disks (both components showing the bar structure), respectively. The stars are selected in the bulge region with $R<5$ kpc and $|Z|<3$ kpc. \label{fig:fig8}}
\end{figure}

\section{\textbf{Auriga Cosmological Simulation}} \label{sec:cossim}
In this section we make use of a cosmological zoom-in simulation of a MW-like galaxy from Auriga to understand the observed pattern in the $V_\phi$-[Fe/H] space and gain insight into the formation and evolution history of the Galaxy.

\begin{figure*}[ht!]
	\plotone{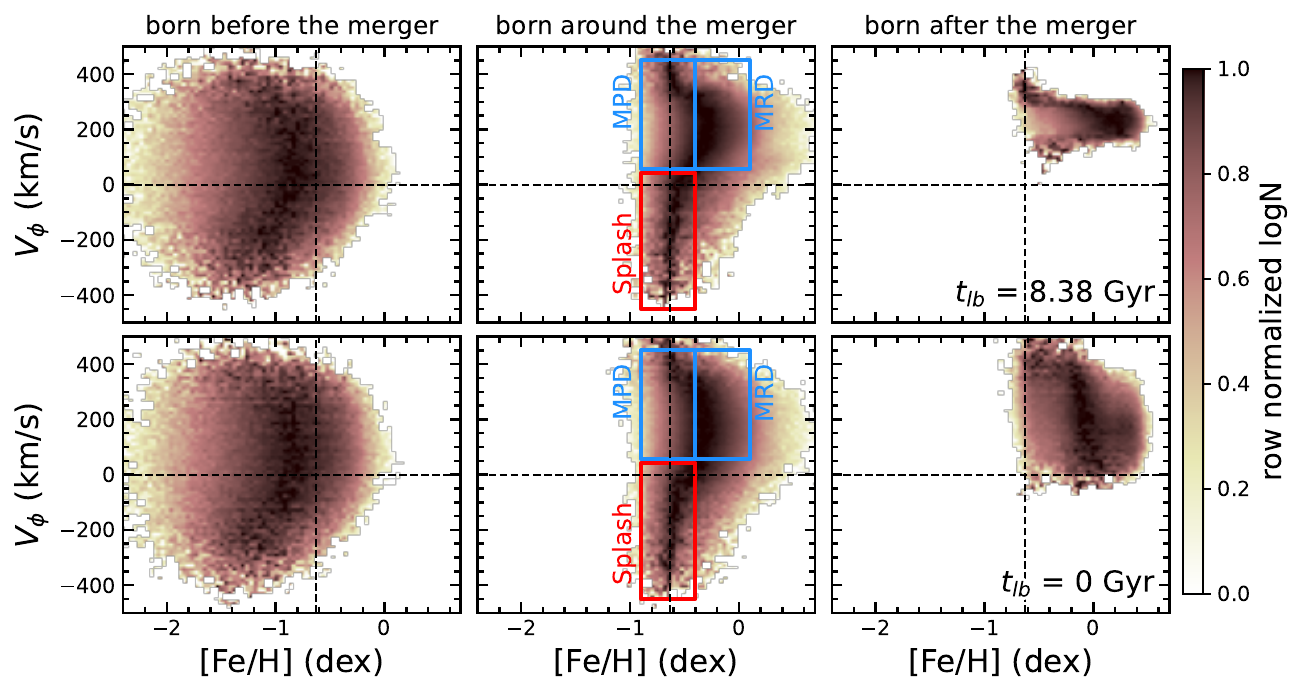}
	\caption{The row-normalized number density maps in the $V_\phi$-[Fe/H] space for stars in the Auriga simulation that were born before the merger (left column), during the merger (middle column), and after the merger (right column). The top and bottom rows show two different look-back time at $t_{lb}=8.38$ Gyr and $t_{lb}=0$ Gyr, respectively. Note the three populations show dramatically different patterns. The three boxes in the middle column represent three subpopulations that are analyzed in detail later. \label{fig:fig9}}
\end{figure*}

Auriga is a suite of 30 magneto-hydrodynamical cosmological zoom-in simulations of MW mass dark matter halos with varying mass from $\sim 1\times10^{12}$ to $2\times10^{12}$ $M_\odot$ and evolving from redshift 127 to the present day, incorporating active galactic nuclei feedback, star formation, magnetic fields (see \citealt{grandAurigaProjectProperties2017} for details). The Auriga simulations \citep{grandOverviewPublicData2024} are publicly available\footnote{\url{https://wwwmpa.mpa-garching.mpg.de/auriga/data}}. We choose Auriga 23 (Au23) which resembles many of the MW properties. For example, it displays clear $[\alpha$/Fe] bimodality for the thin/thick populations \citep{grandOriginChemicallyDistinct2018} and its last major merger event happened at a lookback time $t_{\mathrm{lb}}\approx$10 Gyrs, similar to the GSE merger event \citep{montalbanChronologicallyDatingEarly2021, xiangTimeresolvedPictureOur2022}. Additional information about Au23 regarding its star formation history, bar morphology, rotation curve, and merger history, can be found in \cite{fragkoudiChemodynamicsBarredGalaxies2020a}

\begin{figure*}[ht!]
	\plotone{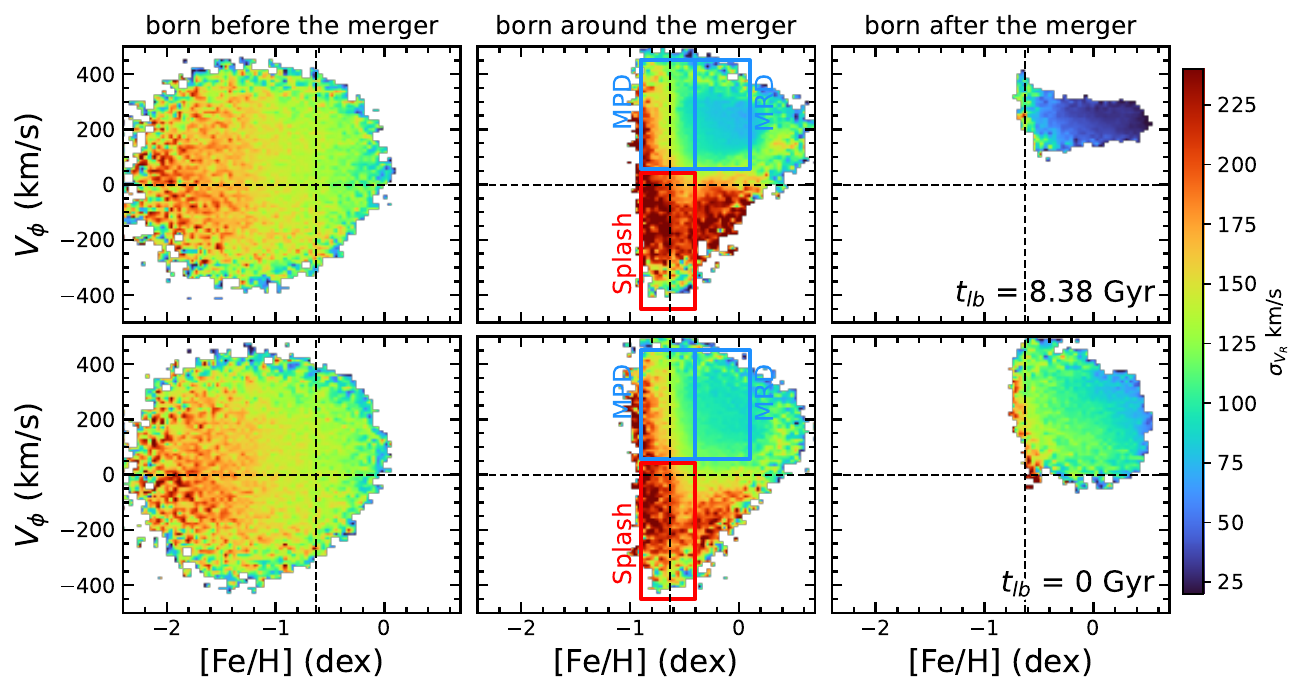}
	\caption{The $V_\phi$-[Fe/H] space color-coded with the radial velocity dispersion for stars born before (left column), during (middle column) and after the merger (right column) in the Auriga simulation Au23. The layout of this figure is the same as Fig.\ref{fig:fig9}. The top and bottom rows represent $t_{lb}=8.38$ Gyr and $t_{lb}=0$ Gyr, respectively.  \label{fig:fig10}}
\end{figure*}

As we have shown in the previous section, chemical information is not enough to distinguish the bulge stellar populations; the addition of $V_\phi$ is crucial to the identification of the non-bar population with negative $V_\phi$ and high radial velocity dispersion in the metallicity range $-1<$[Fe/H]$<-0.5$, which we associate with the Splash. \cite{belokurovBiggestSplash2020} suggested that it is constituted by disk stars heated by the GSE merger in the early epoch. To study the effects of the merger on the formation of the Splash and the bar in the Au23 simulation, we select stars of different ages in the bulge region, dividing them into three epochs: before the last major merger (12 $-$ 13 Gyrs), around the time of the merger (9.6 $-$ 10.4 Gyr) and after the merger (8.5 $-$ 8.9 Gyrs). The last major merger of Au23 happened at around 10 Gyrs (see \citealp{fragkoudiChemodynamicsBarredGalaxies2020a}). Thus, stars of ages at 9.6 $-$ 10.4 Gyr are mainly contributed by the starburst population and dynamically hot disk.

The face-on and edge-on distributions of stars in the inner 5 kpc of Au23 are shown in Fig.~\ref{fig:fig8} (top and bottom row respectively) at $t_{lb} =$ 8.38 Gyr (left column) and 0 Gyr (right column). Stars born before the merger (filled blue contours) resemble an oblate spheroid appearing more circular in the face-on view and elliptical in the edge-on view. The stars born around the merger (black contour lines) and after the merger (red contour lines) contribute to the bar structure. The bar containing stars formed around the merger is slightly shorter and thicker compared to the bar containing stars formed after the merger, consistent with results in literature that stars with hotter kinematics will have thicker and rounder bars than stars with colder kinematics \citep{fragkoudiBarsBoxyPeanut2017a, debattistaSeparationStellarPopulations2017,athanassoulaMetallicitydependentKinematicsMorphology2017}. According to Fig.~\ref{fig:fig8}, in the Au23 simulation, the bar forms quickly after the last major merger when the disk is still dynamically hot. The stars born around and after the merger constitute the thicker and thinner bar, respectively, while the stars born before the merger did not participate in the bar formation process to resemble an oblate structure.

To compare the data (sample B) with the simulation, we narrow down the selection of simulation particles to $2<R<5$ kpc and $|Z|<3$ kpc, to avoid the innermost region with complex kinematics and halo stars. The row-normalized density maps in the $V_\phi -$[Fe/H] space are shown in Fig.~\ref{fig:fig9} for the same lookback times as Fig.~\ref{fig:fig8}. The stars born around the merger (i.e., the middle column) exhibit the chevron-like pattern with a vertical extension towards lower $V_\phi$, very similar to the one in observations (Fig.~\ref{fig:fig5}). In the MW, the top and bottom branches of this chevron-like pattern are attributed to the thin and thick disks, respectively \citep{belokurovDawnTillDisc2022, leeChemodynamicalAnalysisMetalrich2023}. However, as shown in Fig.~\ref{fig:fig9}, in the Auriga simulation, such a chevron-like pattern has already emerged 8.38 Gyr ago, after the last major merger. In the left column of Fig.~\ref{fig:fig8}, the spatial distribution of these stars (black contours) resembles a thick disk with a clear bar structure. This seems to indicate that the chevron-like pattern can be produced even without a significant fraction of thin disk component. The stars born after the merger (top right panel of Fig.~\ref{fig:fig9}) distribute within the upper right region at high metallicities and high $V_\phi$, which is expected since they are born in the disk (see also red contours of Fig.~\ref{fig:fig8}, left column). In Fig.~\ref{fig:fig9} comparing the $V_\phi$-[Fe/H] distributions of stars born before (left column) the merger at $t_{lb}=8.38$ Gyr (top panels) and $t_{lb}=0$ Gyr (bottom panels), there is no clear variation after 8 Gyr of evolution. However, the stars born after the merger (right column) show significant difference in the $V_\phi$-[Fe/H] pattern during the same time period: the pattern shifts to lower azimuthal velocities and has a much larger velocity dispersion. Both the external and internal mechanisms could have contributed to the disk heating. After the last major merger occurred about 10 Gyrs ago, the galaxy experienced subsequent minor mergers (see Fig.~11 in \citealp[]{fragkoudiChemodynamicsBarredGalaxies2020a}) which can also contribute to disk heating. In addition, as shown in Fig.~\ref{fig:fig8}, the stars born after the merger, host a strong bar. The formation of the bar requires loss of angular momentum which leads to larger velocity dispersion.

\begin{figure*}[ht!]
	\includegraphics[width=175mm]{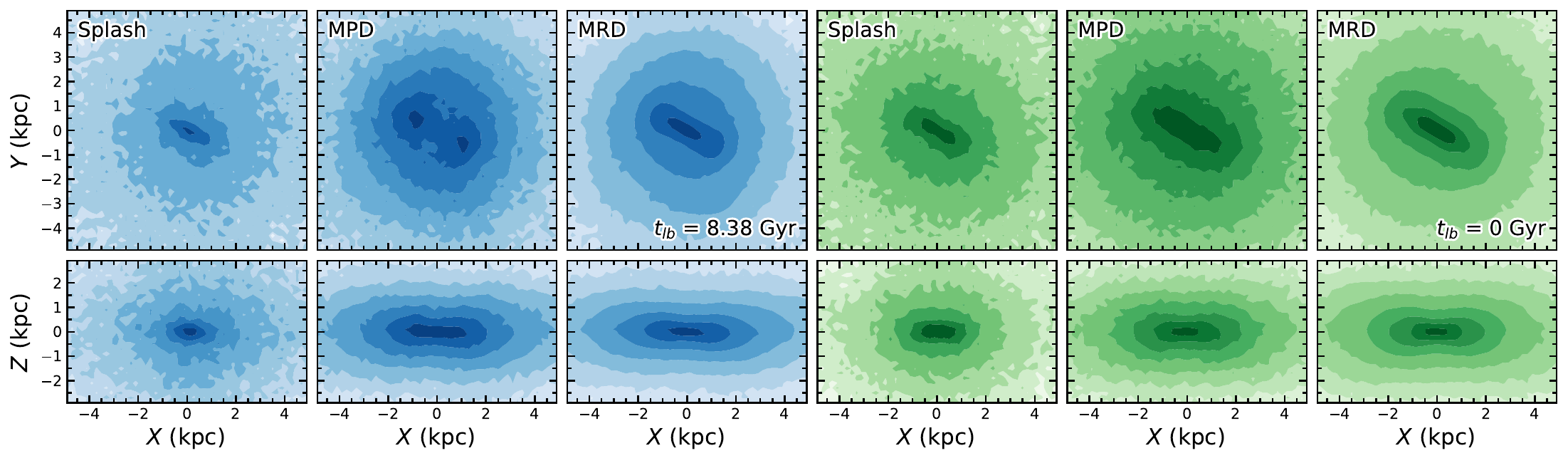}
	\caption{The face-on (top) and edge-on (bottom) images for the Splash, MPD and MRD components at $t_{lb}=8.38$ Gyr (left three columns) and $t_{lb}=0$ Gyr (right three columns). MPD and MRD all show the same bar structure. Splash is overall spherical, with a small bar early on and a relatively weak bar in the late stage.\label{fig:fig11}}
\end{figure*}

\begin{figure*}[ht!] 
	\includegraphics[width=185mm]{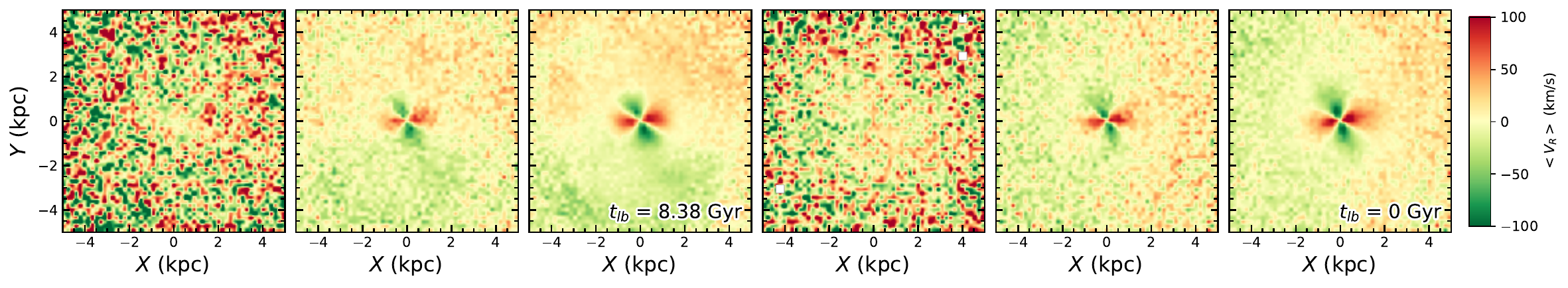}
	\caption{The $\left\langle V_R \right\rangle$ color-coded face-on distribution of Splash, Metal-poor disk (MPD) and Metal-rich disk (MRD) at $t_{lb}=8.38$ Gyr (left three columns) and $t_{lb}=0$ Gyr (right three columns). Consistent with the results in Fig.~\ref{fig:fig7}, both MPD and MRD show the bar-like butterfly pattern, while Splash is random motion dominated. \label{fig:fig12}}
\end{figure*}

We now investigate the $V_\phi -$[Fe/H] space color$-$coded by the radial velocity dispersion $\sigma_{V_R}$ in Fig.~\ref{fig:fig10} and we compare it to the data (bottom panel of Fig.~\ref{fig:fig5}): we notice that stars born around the merger (middle column in Fig.~\ref{fig:fig10}) have a distribution similar to the observations. The velocity dispersion decreases monotonically from lower left corner to the top right corner, consistent with the observed pattern in Fig.~\ref{fig:fig5}. For the stars born before the merger (left column), $\sigma_{V_R}$ is relatively uniform at any given [Fe/H], which is consistent with the behaviour of a dispersion$-$dominated structure.  There is not a  much difference between the $V_\phi -$[Fe/H]$-\sigma_{V_R}$ patterns at $t_{lb}=8$ (top row) and $t_{lb}=0$ (bottom row) for the stars born before (left panels) and around the merger (middle panels), confirming the findings in Fig.~\ref{fig:fig9}: 8 Gyrs of evolution (e.g. the accretion of cold gas, the following disk formation and dynamical heating from secular evolution) do not greatly influence the morphology and kinematics of the stars formed early-on. On the other hand, the younger stars born just after the merger (right column) which are kinematically cold, experience significant heating over the same time period (see bottom right panel).

\begin{figure*}[ht!]
	\plotone{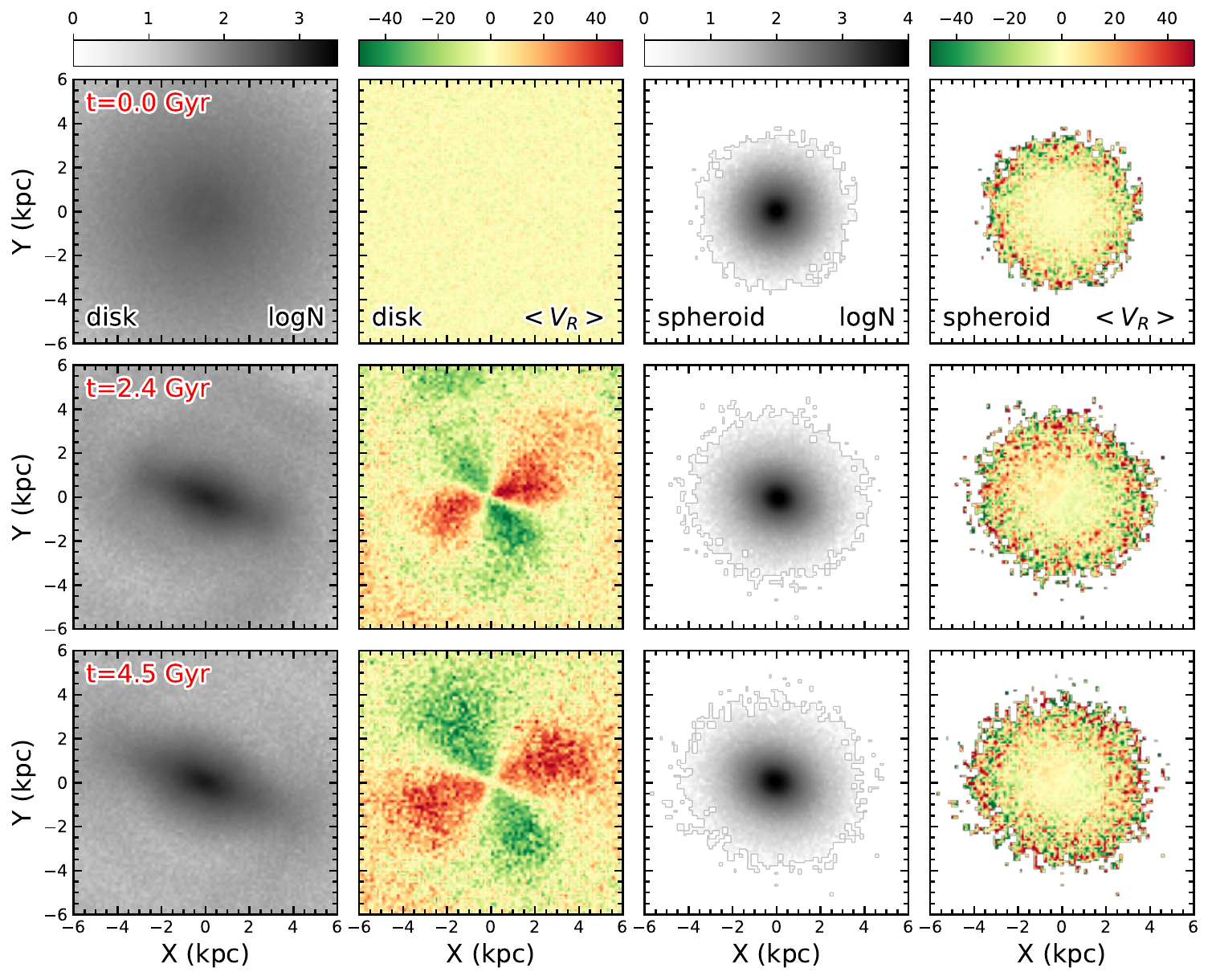}
	\caption{Evolution of morphological and kinematical properties of the disk and the spheroidal component (classical bulge) of the model M1. From top to bottom rows, snapshots at 0.0, 2.4 and 4.5 Gyr are shown respectively. The first two columns are the number density map and median $v_R$ map in $X-Y$ plane for disk particles, while the right two columns are the number density and velocity maps for the spheroidal component. \label{fig:fig13}}
\end{figure*}

We now focus on the stars born around the time of the merger in Au23. We separate them into three sub-populations,MPD (metal-poor disk), MRD (metal-rich disk) and the Splash (see the three boxes in Fig.~\ref{fig:fig9}) to mimic the observed transition from `disk' (top three panels of fourth column in Fig.~\ref{fig:fig7}) to `disk coexisting with Splash' (the middle column in Fig~\ref{fig:fig7}) seen in the bulge Gaia sample. The Splash and MPD have the same metallicity range but the Splash is kinematically hotter with weaker rotation than MPD (see Fig~\ref{fig:fig10}). MPD and MRD have similar $V_\phi$ distributions but MRD is more metal-rich. The face-on and edge-on distributions of these three sub-populations are presented in Fig.~\ref{fig:fig11} at $t_{lb}=8.38$ Gyr (blue iso-density contour maps) and 0 Gyr (green). The corresponding velocity maps are shown in Fig.~\ref{fig:fig12} for the face-on projection. At $t_{lb}=8.38$ Gyr, after the last major merger when the bar just formed, the Splash is nearly spherical with the inner 1 kpc showing a bar-like morphology. MPD and MRD both contain a bar structure within 2 kpc from the GC, with the bar in MPD being slightly thicker than the bar in MRD. However at redshift 0, both MPD and MRD have evolved to have almost the same morphology, including a similar-size bar/bulge. The Splash component also develops a weak bar ($\sim2$ kpc half length) following the bar orientation in MPD and MRD. However, as shown in the $\left\langle V_R \right\rangle$ color-coded maps in Fig.~\ref{fig:fig12}, the Splash does not display the typical bar-like butterfly pattern seen in MPD and MRD with a very prominent bar in the inner 2 kpc -- this is consistent with the data shown in Fig~\ref{fig:fig7}. Therefore, the Splash stars likely do not contribute to the bar supporting orbits. The bar-like appearance of the Splash in the inner 2 kpc at $t_{lb}=0$ is likely due to the gravitational attraction from the bar. Nonetheless, by looking at the mean $V_R$ maps alone, we cannot exclude the possibility that a few of the Splash stars indeed move on bar supporting orbits. Beyond 2 kpc, the Splash morphology becomes more spheroidal, resembling a `fluffy classical bulge' (see Fig. 11 in \citealt{belokurovBiggestSplash2020}). 


\section{\textbf{ Interaction between the Bar and a Pre-existing Bulge: $N$-body Simulation Test}} \label{sec:nbody}

\begin{figure*}[ht!]
	\plotone{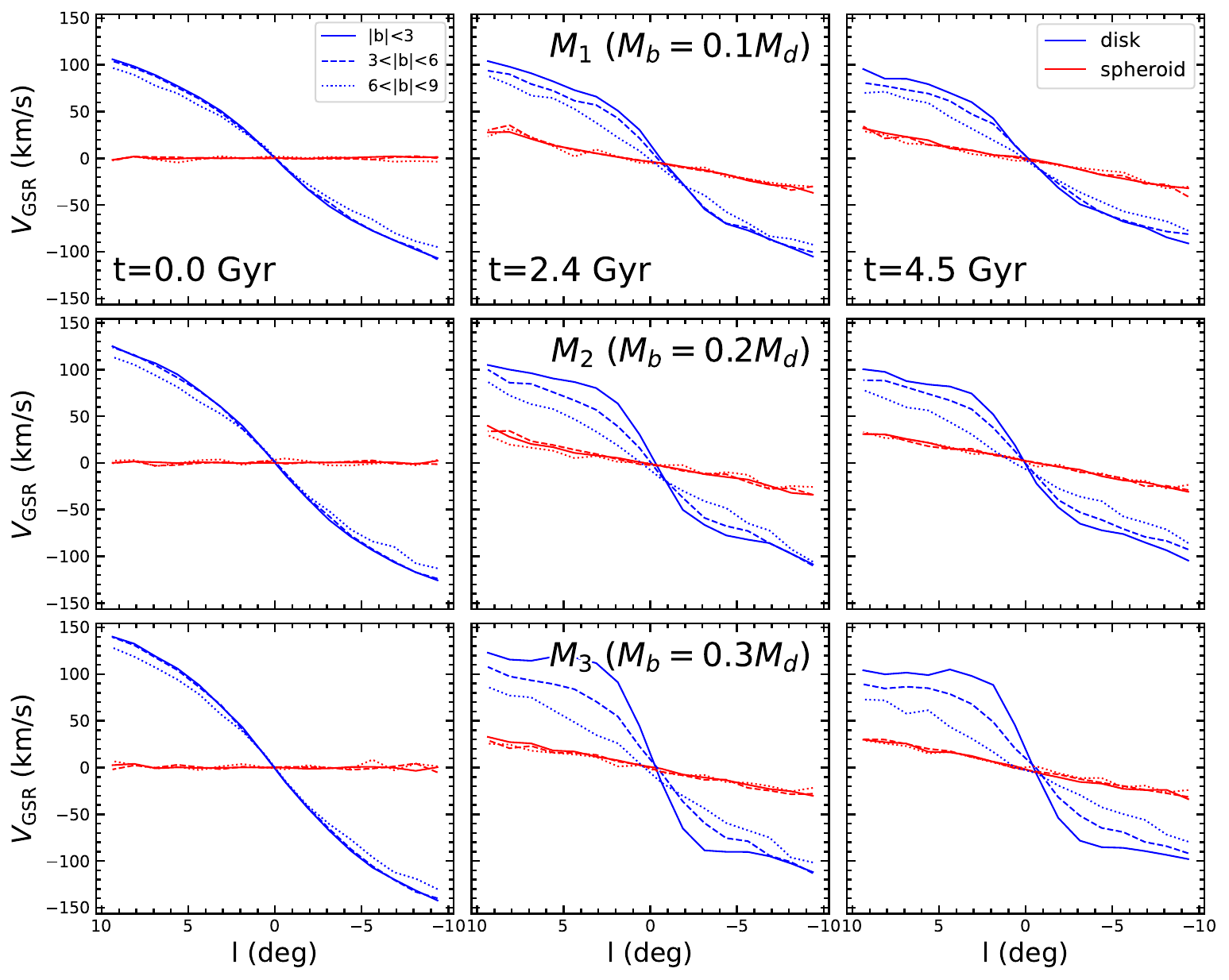}
	\caption{Evolution of the rotation profiles of the disk and spheroidal components in the mock observation of the bulge region for the three models. The blue and red curves represent the disk (bar) and bulge rotation profiles, respectively. The mass fraction of the pre-existing classical bulge increases from top to bottom panels. Apparently the classical bulge would be spun up by the bar. And in more massive bulge, the bar deviate from the perfect the cylindrical rotation. \label{fig:fig14}}
\end{figure*}

To investigate the influence of a rotating bar on a pre-existing non-rotating spheroidal structure (i.e., a classical bulge) and the origin of the observed cylindrical rotation for the metal-poor stars in the Galactic bulge region, we make use of three isolated $N$-body simulations of MW-like galaxies that self-develop a bar out of disk particles. We set up the initial condition (IC) following \cite{tepper-garciaBarredMilkyWay2021}, with a dark matter halo ($M_h\sim 1.2\times10^{12}$ $M_\odot$), a stellar disk ($M_d= 4.3\times10^{10}$ $M_\odot$) with a scale length of 2.5 kpc and a scale height of 0.3 kpc (resembling a thin disk) and a non-rotating stellar spheroidal structure - the pre-existing classical bulge. We build three models, namely, M1, M2 and M3, each with different classical bulge mass $M_b=0.1M_d, 0.2M_d$ and $0.3 M_d$. The dark matter halo, disk, and bulge in all models are approximated by $10^6$, $10^6$, and $6\times10^5$ particles, respectively. More simulation details and the final resemblance with the MW can be found in \cite{tepper-garciaBarredMilkyWay2021}. The ICs are generated using iterative self-consistent modelling module in AGAMA \citep{vasilievAGAMAActionbasedGalaxy2019}, which are then evolved with GADGET4 \citep{springelSimulatingCosmicStructure2021} for 5 Gyrs. A bar is fully formed and buckles into a peanut-shaped bulge within $\sim$1.5 Gyr for M1, the model with the smallest bulge mass. The formation time of the bar is delayed for M2 and M3 with larger bulge mass. This is consistent with the findings of \cite{liEvolutionStellarBars2023}.

\begin{figure*}[ht!]
	\plotone{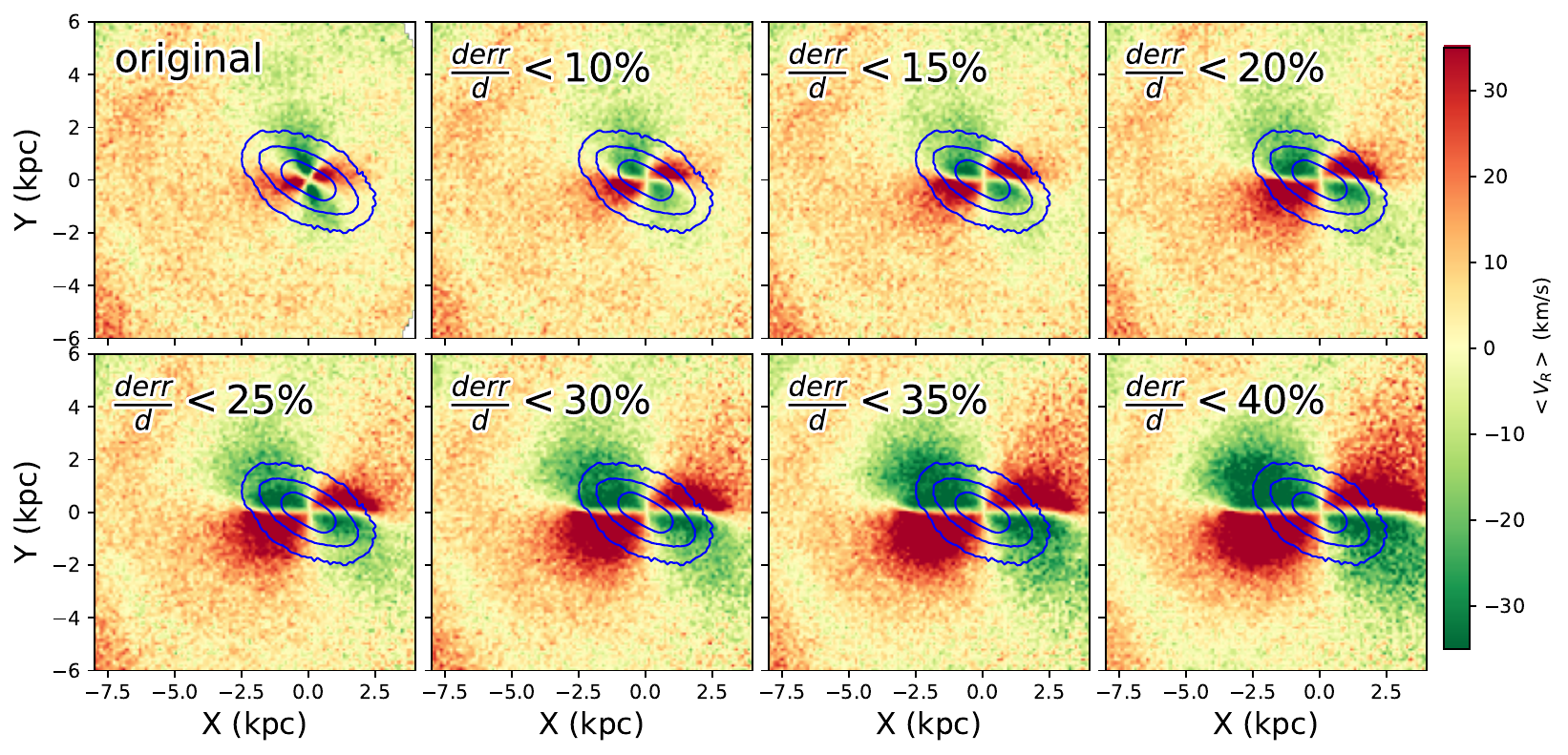}
	\caption{Effect of the relative distance error ($d_{err}/d$) on the mean $\left\langle V_R \right\rangle$ map in $X-Y$ plane of the Au23 simulation at $t_{lb}=0$ Gyr. The top left panel shows the original map and the other panels are the perturbed maps with different levels of distance uncertainty as indicated by the text at the upper left corner of each panel. The contour in each panel is the density contour of the original bar structure. With the increasing distance uncertainties, the $\left\langle V_R \right\rangle$ butterfly pattern becomes more significant, with the zero velocity line more closely aligned with the $X$-axis rather than the bar major axis ($20^\circ$ tilt from the $X$-axis). \label{fig:fig15}}
\end{figure*}

The temporal evolution of the morphology and kinematics of M1 is shown in Fig.~\ref{fig:fig13}, for the disk (left two columns) and non-rotating spheroidal component (right two columns). The number density and mean $V_R$ maps in the $X-Y$ plane reveal that the disk particles at 0 Gyrs (top row) have an axisymmetric distribution without any velocity pattern. Once a bar forms (middle row), the butterfly pattern appears in the $\left\langle V_R \right\rangle$ map due to the systematic motion of the bar orbits in different quadrants. As the bar grows, the $\left\langle V_R \right\rangle$  butterfly pattern becomes larger (bottom row). On the other hand, the spheroidal non-rotating component does not show the $\left\langle V_R \right\rangle$  butterfly pattern even at the end of the simulation, indicating that the pre-existing spheroidal bulge component is not involved in the bar formation. Instead, it morphs into a slightly elliptical structure aligning with the bar major axis, due to the gravitational attraction from the bar. 

To further investigate the evolution of the bulge in a more quantitative way, we generate mock observations of the Galactic bulge from the simulations by setting the observer at $R=8$ kpc and a bar viewing angle of $20^\circ$. The longitudinal rotational profiles of the particles are shown in Fig.~\ref{fig:fig14} for the disk (blue curves) and spheroidal non-rotating component (red curves) at different times (increasing from left to right columns). M1, M2 and M3 are shown in the top, middle, and bottom rows, respectively. The rotation profiles of the non-rotating component are initially flat, with no net rotation, but become steeper with time as it gains angular momentum from the rotating bar \citep{sahaSpinupLowmassClassical2012} for all three models; its morphology also changes (see Fig~\ref{fig:fig13}) from spheroidal to ellipsoidal, with the major axis aligned with the bar major axis. In Fig.~\ref{fig:fig13}, we have seen that classical bulges may not display the $\left\langle V_R \right\rangle$ butterfly pattern, suggesting that they are not building blocks for the bar. However, the classical bulges in all three models evolve to exhibit cylindrical rotation, which is a known property of bars and boxy/peanut-shaped bulges in $N$-body simulations \citep{athanassoulaMorphologyPhotometryKinematics2002, sahaSecularEvolutionCylindrical2013}. This work confirms the result of \cite{sahaSpinupLowmassClassical2012} that cylindrical rotation is not an unique property of bars, e.g. a pre-existing classical bulge can be spun up by the bar to result in cylindrical rotation. However, the disk particles (the bar) do not strictly follow the cylindrical rotation. Although the rotational velocities at different latitude bins almost converge at $|l|\sim10^\circ$, the rotation curves in inner region change with latitude and the discrepancy is more significant for models with larger classical bulge mass. In the MW, the rotation curves in the bulge region barely change with latitude (\citealp[also see Fig.~\ref{fig:fig6}]{howardKINEMATICSEDGEGALACTIC2009,kunderBULGERADIALVELOCITY2012,zoccaliGIRAFFEInnerBulge2014}). This result could put constrains on the mass of the MW classical bulge, which should be less than 10\% of the disk mass, consistent with \citet{shenOURMILKYWAY2010a}.

Although our simulations are pure $N$-body simulations with no chemical information, the initially non-rotating spheroidal component with high velocity dispersion could be considered a classical bulge, formed in the early chaotic stages of the MW. In observations, it would likely correspond to the (old) metal-poor population ([Fe/H]$\lesssim-1$ dex). In our bulge sample B, stars with $-1.5\lesssim $[Fe/H] $\lesssim-1$ dex do not exhibit the  $\left\langle V_R\right\rangle$ butterfly-pattern (Fig.~\ref{fig:fig7}) but show cylindrical rotation (Fig.~\ref{fig:fig6}), while the more metal-poor stars ($-2<$[Fe/H]$<-1.5$ dex) do not display cylindrical rotation or the butterfly-pattern. Our interpretation is that the more metal-rich stars ($-1.5\lesssim [Fe/H] \lesssim-1$ dex) could represent classical bulge spun up by the bar while the more metal-poor ones ($-2<$[Fe/H]$<-1.5$ dex) could be halo stars, which spend more time outside of the bar's influence. Therefore, their kinematics is less easily affected by the bar presence.   

\section{Discussion} \label{sec:dis}

\subsection{Effect of Distance Error on the $\left\langle V_R \right\rangle$ Map}

Bulge stars have large distance measurement uncertainties due to the their larger distance, strong dust extinction and crowding. To test the influence of the distance uncertainty on the $\left\langle V_R \right\rangle$ maps projected onto the $X-Y$ plane (i.e., the butterfly pattern), we first transform the Cartesian coordinates $(X, Y, Z, V_X, V_Y, V_Z)$ of the simulation into observables $(ra, dec, d, pmra, pmdec, rv)$ then perturb the heliocentric distance $d$ with various levels of uncertainty, and finally transform the observational quantities into Galactic cylindrical coordinates. Fig.~\ref{fig:fig15} shows the original $\left\langle V_R \right\rangle$ map (top left panel) and distance-perturbed maps with the corresponding relative distance uncertainty ($d_{err}$/$d$) listed in the upper left corner. The butterfly pattern becomes more significant as the relative uncertainties increase and the zero-velocity line gradually aligns with the Sun-GC line (the $X$-axis) as they reach $d_{err}$/$d$ $\approx$25\%. Beyond 25 \%, the zero-velocity line remains aligned with the $X$-axis and the butterfly pattern keeps becoming more significant. The $\left\langle V_R \right\rangle$ butterfly-pattern is therefore a robust bar signature: in the case of bar presence, large distance uncertainties will not cause the pattern to disappear, the pattern will merely twist such that the zero-velocity line becomes more aligned with the Sun-GC line-of-sight (see also \citealp[]{visloskyGaiaDR3Data2024}).

\subsection{Splash Formation Scenario and the Early Formation of the Disk}

The Splash in the $V_\phi$-[Fe/H] space appears as a smooth transition from thick disk to a dynamically hot component with small or retrograde rotation velocity (Figs.~\ref{fig:fig3} and~\ref{fig:fig4}). A natural explanation for the formation of the Splash is that it consists of thick disk stars heated by a major merger \citep{belokurovBiggestSplash2020}. The Auriga cosmological simulation \citep{grandDualOriginGalactic2020} also contains a Splash-like population. This scenario assumes that a thick disk was already in place before the major merger. Another mechanism that could be responsible for the formation of the Splash is the clumpy formation scenario; clumps form in the disk due to the gas fragmentation in the early gas-rich stage \citep{clarkeImprintClumpFormation2019a}. The clumps have high star formation rate density to self-enrich in $\alpha$ elements rapidly, which are then destroyed by supernovae to dissolve into a high-$\alpha$ thick disk. Meanwhile, across the disk, the continuous long term star formation gradually gives rise to a low-$\alpha$ thin disk. In the clumpy formation scenario it is possible to form a Splash-like population, which corresponds to the low angular momentum tail of the thick disk \citep{amaranteSplashMerger2020}. 

We have shown in Section \ref{subsec:dtob}, Fig.~\ref{fig:fig3},~\ref{fig:fig4} that the Splash exists across the disk as an uniform population with similar metallicities at different Galactic radii, which can be explained by both scenarios above. In the first scenario, there already was a thick disk in place at the time of the last major merger, which had formed in the early chaotic epoch of the galaxy. The chemistry of such a disk would be spatially well-mixed, implying that stars kicked out of the thick disk into halo-like orbits by the major merger would have similar metallicities at different radii. In the clumpy formation mechanism, the proto-thick-disk originates from the clumps formed in the first Gyrs during the gas-rich phase. These clumps were formed out of the gas with similar chemistry, experiencing a similar chemical enrichment process. In this scenario, the metallicity of the thick disk is also well-mixed. In addition, \cite{leeChemodynamicalAnalysisMetalrich2023} reported that there is a relatively more metal-poor and $\alpha$-depleted component in the Splash region. The author attributed approximately half of the stars in this component to the accreted GSE and the other half to the starburst population formed during the merger. However, the low-$\alpha$ Splash component can also be explained by the clumpy formation scenario if the early clumps scattered thin disk stars with low-$\alpha$ abundances to hotter orbits. The current observations of the Splash stars can not discriminate between the two scenarios. It is also possible that both scenarios could have played a role in the formation of the Splash.

\subsection{Why is [Fe/H] $\sim-1$ Special?}

The multiple stellar populations and structures in the Galactic bulge make its chemodynamics complex. Most Galactic bulge studies have used the metallicity as a rough tracer of different stellar populations to explore its structure and kinematics. Many surveys have revealed that the stellar populations with [Fe/H] $\gtrsim-1$ dex show bar-like kinematics (\citealp[i.e., have disk origin]{kunderBULGERADIALVELOCITY2012,nessARGOSIVKinematics2013, dimatteoDiscOriginMilky2016a, fragkoudiDiscOriginMilky2018}), while stars with [Fe/H] $\lesssim-1$ dex are consistent with a dispersion-dominant sphroidal structure \citep{arentsenPristineInnerGalaxy2020,dekanyVVVSurveyNearinfrared2013}. An important question naturally arises: what is the reason for this transition at [Fe/H] $=-1$ dex? 

As shown in Fig.~\ref{fig:fig3}, the metallicity distribution of the Splash at different radii is within $-1<$ [Fe/H] $<-0.4$ dex; the lower metallicity limit of the Splash is approximately [Fe/H] $\sim-1$ dex, where the kinematics changes from bar-like to dispersion-dominated. Under the assumption that the formation of the Splash is merger-induced and mainly composed of heated proto-disk stars and the subsequent starburst population that is more metal-rich than the proto-disk (with partly overlapping metallicities), then the most metal-poor stars in the Splash cannot be more metal-poor than the proto-disk itself. Therefore, the [Fe/H] $\sim-1$ dex limit should be the lower limit of the proto-disk at the time of the major merger event. Afterwards, the bar forms out of the dynamically hot disk stars. This is consistent with recent simulations showing that a bar can form from the thick disk \citep{ghoshBarsBoxyPeanut2023}. Thus, the disk stars with [Fe/H] $\gtrsim-1$ dex become the building blocks of the bar and kinematically detach from the more metal-poor stars. However, most of the hotter Splash population with [Fe/H] $\gtrsim-1$ dex does not participate in the formation of the bar; instead, the kinematics of the Splash is similar to that of stars with [Fe/H] $<-1$ dex. However, in Section 4.1 we have shown that the morphology of the Splash could become bar-like in the inner 2 kpc, which could be due to the long-term gravitational attraction of the bar. Beyond 2 kpc, the Splash becomes more spheroidal like (e.g., \citealt{belokurovBiggestSplash2020}).

In addition, other works also suggest that [Fe/H] $\sim-1$ dex is a turning point in the galaxy chemical evolution. Based on the Gaia data with the metallicity derived from XGBoost, \cite{rixPoorOldHeart2022} reported that the slope of the metallicity distribution function is $d(\log n_*)/d{\rm [M/H]}\sim1$ in the bulge region for the bulk of stars with [Fe/H] $\lesssim-1$. In the one-zone chemical evolution model \citep{weinbergEquilibriumSuddenEvents2017}, $d(\log n_*)/d{\rm [M/H]}\sim1$ is a natural consequence of a self-enriching in-situ system. The slope then becomes steeper at [Fe/H] $\gtrsim -1$ dex, which implies an event with rapid gas accretion (or merger), corresponding to the onset of the thick disk \citep{belokurovDawnTillDisc2022,semenovFormationGalacticDisks2023}. \cite{chandraThreePhaseEvolutionMilky2023} also used red giants with Gaia XP spectra to study the  Milky Way evolution in a similar parameter space ($L_z/L_c$-[Fe/H]) as this work ($V_\phi$-[Fe/H]). They found that the old high-$\alpha$ disk starts at [Fe/H] $\sim -1$ dex, which agrees well with our results on the proto-disk, for which we estimate a lower metallicity limit of [Fe/H] $\sim -1$ dex.

Using a large sample of subgiants with precise age measurements, \cite{xiangTimeresolvedPictureOur2022} separated the sample into early phase (low angular momentum and high $\alpha$, i.e., thick disk and halo) and late phase (high angular momentum and low $\alpha$, i.e., thin disk) stars; thick disk stars in the early phase have [Fe/H] $\gtrsim-1$ dex. This metallicity boundary agrees with our inference from the Galactic bulge observations that the low metallicity limit of the thick disk is [Fe/H] $\approx-1$ dex.

\section{Conclusion} \label{sec:con}

Recent Solar neighbourhood studies have found that the thin disk, thick disk and the Splash appear as distinct features of the $V_\phi$-[Fe/H] diagram \citep{belokurovBiggestSplash2020, leeChemodynamicalAnalysisMetalrich2023}. In this work, we used data from \cite{andraeRobustDatadrivenMetallicities2023}, a catalog with metallicities extracted from the Gaia-XP spectra, to study the $V_\phi$-[Fe/H] diagram over a large range of radii, from the Galactic bulge to the outer disk, up to radius of 14 kpc. We find that the characteristic patterns of the disks and Spash in the $V_\phi$-[Fe/H], exist across the disk. The $V_\phi$-[Fe/H] profiles of the thin disk show systematic variation with radius, which is in agreement with the inside-out formation scenario. The $V_\phi$-[Fe/H] profiles at various radii for the thick disk and the Splash are quite consistent, implying their early formation timescale when the chemistry was spatially well mixed.

The bulge stars share a similar pattern in the $V_\phi$-[Fe/H] space with disk stars, implying that it originated from the disk. By investigating the bar signatures of stellar populations in $V_\phi$-[Fe/H] space and only considering the populations with reliable number statistics, we found that:
\begin{itemize}
  \item Stellar populations with [Fe/H] $\lesssim-1$ dex are dispersion-dominanted.
  \item Stellar populations with $-1\lesssim$ [Fe/H] $\lesssim -0.4$, show bimodal distribution. The stars with $V_{\phi}\gtrsim 100$ $\rm km\ s^{-1}$ follow the bar kinematics while the stars with $V_{\phi}\lesssim 100$ $\rm km\ s^{-1}$ (Splash) have dispersion dominated kinematics similar to those more metal-poor populations.
  \item Stellar populations with [Fe/H] $\gtrsim -0.4$ dex all have bar-like kinematics.
\end{itemize}
By analyzing a MW-like Auriga simulation Au23, we found that only the stars born around the time of the last major merger ($\sim 10$ Gyr ago), which is mainly contributed by the starburst population and the dynamically hot disk, have a $V_\phi$-[Fe/H] pattern similar to the observation in the Galactic bulge. A possible picture for the Galactic evolution emerges based on the great resemblance between the simulation and observations, that is, a preliminary thick disk was already in place not long before the last major merger. Afterwards, the occurrence of the last major merger heated some of the thick disk stars and triggered starburst formation that contribute to the Splash, whose lower metallicity limit is $\sim-1$ dex that is inherited from the thick disk. Thus the lower metallicity limit of the thick disk should be $\sim-1$ dex at the time of the last major merger. Subsequently, the thick disk stars with [Fe/H] $\gtrsim-1$ dex form the bar structure, their kinematics becomes bar-like and detach from those more metal-poor stars. In another aspect, the merger kicks some of thick disk stars to halo-like orbits and induces starburst formation, mainly giving rise to the Splash population, which does not participate in the bar formation. Thus its kinematics is more dispersion-dominated.

Moreover, the observed metal-poor stars ($-1.5<$ [Fe/H] $<-1$ dex) show cylindrical rotation without butterfly pattern in their mean $V_R$ map, possibly as a consequence of the bar losing angular momentum to a pre-existing classical-bulge like structure \citep{sahaSpinupLowmassClassical2012} during secular evolution. We also perform three $N$-body simulations to study the interaction between an initially non-rotating spheroidal component (a classical bulge) and a later formed bar and confirm the result of \cite{sahaSpinupLowmassClassical2012} that an initially non-rotating spheroidal component can be spun up to develop cylindrical rotation under the bar influence without following the bar orbits. In addition, we found that the bulge mass affects the characteristic rotation profiles of the bar. The three $N$-body models are initiated with different bulge masses. Although the rotation profiles ($l-V_{gsr}$) of the bars in the three models almost converge at $l\sim10^\circ$ for the different Galactic latitudes we consider, in the inner region they show relatively large discrepancies:
\begin{itemize}
  \item For the model with the least massive spheroid component ($M_1$), the rotation profiles of the bar at different latitudes are similar to each other, consistent with the cylindrical rotation pattern.
  \item For the model with the most massive spheroid component ($M_3$), the rotation profiles of the bar at higher latitudes have much shallower slopes at $|l| \lesssim 4^{\circ}$ compared to the lower latitudes.
\end{itemize}
In the Milky Way, the rotation profiles at different latitudes are almost the same (see Fig.~\ref{fig:fig6}), which is most close to the model with the least massive bulge, implying that the classical bulge in MW, if there is one, should not be too large (less than 10\% disk mass). In the future, we plan to consider other parameters of the classical bulge such as spin, velocity dispersion, and velocity anisotropy, etc., to quantitatively match to observations and further constrain the mass and other properties of the MW classical bulge.

\begin{acknowledgments}
    This work is supported by the National Natural Science Foundation of China under grant No. 12122301, 12233001, by a Shanghai Natural Science Research Grant (21ZR1430600), by the ``111'' project of the Ministry of Education under grant No. B20019, and by the China Manned Space Project with No. CMS-CSST-2021-B03. We thank the sponsorship from Yangyang Development Fund. ITS thanks for the support of the Shanghai Key Lab for Astrophysics and the National Natural Science Foundation of China, grant no. 12203035. SRG is supported by an STFC Ernest Rutherford Fellowship (ST/W003643/1). This work made use of the Gravity Supercomputer at the Department of Astronomy, Shanghai Jiao Tong University.
\end{acknowledgments}

\bibliography{ref}{}

\begin{thebibliography}{}
\expandafter\ifx\csname natexlab\endcsname\relax\def\natexlab#1{#1}\fi
\providecommand{\url}[1]{\href{#1}{#1}}
\providecommand{\dodoi}[1]{doi:~\href{http://doi.org/#1}{\nolinkurl{#1}}}
\providecommand{\doeprint}[1]{\href{http://ascl.net/#1}{\nolinkurl{http://ascl.net/#1}}}
\providecommand{\doarXiv}[1]{\href{https://arxiv.org/abs/#1}{\nolinkurl{https://arxiv.org/abs/#1}}}

\bibitem[{Amarante {et~al.}(2020)Amarante, Beraldo E~Silva, Debattista, \& Smith}]{amaranteSplashMerger2020}
Amarante, J. A.~S., Beraldo E~Silva, L., Debattista, V.~P., \& Smith, M.~C. 2020, ApJ, 891, L30, \dodoi{10.3847/2041-8213/ab78a4}

\bibitem[{Andrae {et~al.}(2023)Andrae, Rix, \& Chandra}]{andraeRobustDatadrivenMetallicities2023}
Andrae, R., Rix, H.-W., \& Chandra, V. 2023, Robust {{Data-driven Metallicities}} for 120 {{Million Stars}} from {{Gaia XP Spectra}},  arXiv, \dodoi{10.48550/arXiv.2302.02611}

\bibitem[{Antoja {et~al.}(2018)Antoja, Helmi, {Romero-G{\'o}mez}, Katz, Babusiaux, Drimmel, Evans, Figueras, Poggio, Reyl{\'e}, Robin, Seabroke, \& Soubiran}]{antojaDynamicallyYoungPerturbed2018}
Antoja, T., Helmi, A., {Romero-G{\'o}mez}, M., {et~al.} 2018, Nature, 561, 360, \dodoi{10.1038/s41586-018-0510-7}

\bibitem[{Arentsen {et~al.}(2020)Arentsen, Starkenburg, Martin, Hill, Ibata, Kunder, Schultheis, Venn, Zucker, Aguado, Carlberg, Gonz{\'a}lez~Hern{\'a}ndez, Lardo, Longeard, Malhan, Navarro, {S{\'a}nchez-Janssen}, Sestito, Thomas, Youakim, Lewis, Simpson, \& Wan}]{arentsenPristineInnerGalaxy2020}
Arentsen, A., Starkenburg, E., Martin, N.~F., {et~al.} 2020, Monthly Notices of the Royal Astronomical Society: Letters, 491, L11, \dodoi{10.1093/mnrasl/slz156}

\bibitem[{Athanassoula \& Misiriotis(2002)}]{athanassoulaMorphologyPhotometryKinematics2002}
Athanassoula, E., \& Misiriotis, A. 2002, Monthly Notices of the Royal Astronomical Society, 330, 35, \dodoi{10.1046/j.1365-8711.2002.05028.x}

\bibitem[{Athanassoula {et~al.}(2017)Athanassoula, Rodionov, \& Prantzos}]{athanassoulaMetallicitydependentKinematicsMorphology2017}
Athanassoula, E., Rodionov, S.~A., \& Prantzos, N. 2017, Mon. Not. R. Astron. Soc: Lett., 467, L46, \dodoi{10.1093/mnrasl/slw255}

\bibitem[{{Bailer-Jones} {et~al.}(2021){Bailer-Jones}, Rybizki, Fouesneau, Demleitner, \& Andrae}]{bailer-jonesEstimatingDistancesParallaxes2021}
{Bailer-Jones}, C. A.~L., Rybizki, J., Fouesneau, M., Demleitner, M., \& Andrae, R. 2021, The Astronomical Journal, 161, 147, \dodoi{10.3847/1538-3881/abd806}

\bibitem[{Barbuy {et~al.}(2018)Barbuy, Chiappini, \& Gerhard}]{barbuyChemodynamicalHistoryGalactic2018}
Barbuy, B., Chiappini, C., \& Gerhard, O. 2018, Annual Review of Astronomy and Astrophysics, 56, 223, \dodoi{10.1146/annurev-astro-081817-051826}

\bibitem[{Belokurov {et~al.}(2018)Belokurov, Erkal, Evans, Koposov, \& Deason}]{2018MNRAS.478..611B}
Belokurov, V., Erkal, D., Evans, N.~W., Koposov, S.~E., \& Deason, A.~J. 2018, Monthly Notices of the Royal Astronomical Society, 478, 611, \dodoi{10.1093/mnras/sty982}

\bibitem[{Belokurov \& Kravtsov(2022)}]{belokurovDawnTillDisc2022}
Belokurov, V., \& Kravtsov, A. 2022, Monthly Notices of the Royal Astronomical Society, 514, 689, \dodoi{10.1093/mnras/stac1267}

\bibitem[{Belokurov {et~al.}(2020)Belokurov, Sanders, Fattahi, Smith, Deason, Evans, \& Grand}]{belokurovBiggestSplash2020}
Belokurov, V., Sanders, J.~L., Fattahi, A., {et~al.} 2020, Monthly Notices of the Royal Astronomical Society, 494, 3880, \dodoi{10.1093/mnras/staa876}

\bibitem[{Binney {et~al.}(1991)Binney, Gerhard, Stark, Bally, \& Uchida}]{binneyUnderstandingKinematicsGalactic1991}
Binney, J., Gerhard, O.~E., Stark, A.~A., Bally, J., \& Uchida, K.~I. 1991, Monthly Notices of the Royal Astronomical Society, 252, 210, \dodoi{10.1093/mnras/252.2.210}

\bibitem[{{Bland-Hawthorn} {et~al.}(2023){Bland-Hawthorn}, {Tepper-Garcia}, Agertz, \& Freeman}]{bland-hawthornRapidOnsetStellar2023}
{Bland-Hawthorn}, J., {Tepper-Garcia}, T., Agertz, O., \& Freeman, K. 2023, ApJ, 947, 80, \dodoi{10.3847/1538-4357/acc469}

\bibitem[{Bovy {et~al.}(2019)Bovy, Leung, Hunt, Mackereth, {Garc{\'i}a-Hern{\'a}ndez}, \& {Roman-Lopes}}]{bovyLifeFastLane2019b}
Bovy, J., Leung, H.~W., Hunt, J. A.~S., {et~al.} 2019, Monthly Notices of the Royal Astronomical Society, 490, 4740, \dodoi{10.1093/mnras/stz2891}

\bibitem[{Chandra {et~al.}(2023)Chandra, Semenov, Rix, Conroy, Bonaca, Naidu, Andrae, Li, \& Hernquist}]{chandraThreePhaseEvolutionMilky2023}
Chandra, V., Semenov, V.~A., Rix, H.-W., {et~al.} 2023, The {{Three-Phase Evolution}} of the {{Milky Way}},  arXiv.
\newblock \doeprint{2310.13050}

\bibitem[{Clarke {et~al.}(2019)Clarke, Debattista, Nidever, Loebman, Simons, Kassin, Du, Ness, Fisher, Quinn, Wadsley, Freeman, \& Popescu}]{clarkeImprintClumpFormation2019a}
Clarke, A.~J., Debattista, V.~P., Nidever, D.~L., {et~al.} 2019, Monthly Notices of the Royal Astronomical Society, 484, 3476, \dodoi{10.1093/mnras/stz104}

\bibitem[{Clarkson {et~al.}(2018)Clarkson, Calamida, Sahu, Brown, Gennaro, Avila, Valenti, Debattista, Rich, Minniti, Zoccali, \& Aufdemberge}]{clarksonChemicallyDissectedRotation2018}
Clarkson, W.~I., Calamida, A., Sahu, K.~C., {et~al.} 2018, The Astrophysical Journal, 858, 46, \dodoi{10.3847/1538-4357/aaba7f}

\bibitem[{Collaboration {et~al.}(2022{\natexlab{a}})Collaboration, Vallenari, Brown, Prusti, {de Bruijne}, Arenou, Babusiaux, Biermann, Creevey, Ducourant, Evans, Eyer, Guerra, Hutton, Jordi, Klioner, Lammers, Lindegren, Luri, Mignard, Panem, Pourbaix, Randich, Sartoretti, Soubiran, Tanga, Walton, {Bailer-Jones}, Bastian, Drimmel, Jansen, Katz, Lattanzi, {van Leeuwen}, Bakker, Cacciari, Casta{\~n}eda, De~Angeli, Fabricius, Fouesneau, Fr{\'e}mat, Galluccio, Guerrier, Heiter, Masana, Messineo, Mowlavi, Nicolas, Nienartowicz, Pailler, Panuzzo, Riclet, Roux, Seabroke, Sordo{\o}rcit, Th{\'e}venin, {Gracia-Abril}, Portell, Teyssier, Altmann, Andrae, Audard, {Bellas-Velidis}, Benson, Berthier, Blomme, Burgess, Busonero, Busso, C{\'a}novas, Carry, Cellino, Cheek, Clementini, Damerdji, Davidson, {de Teodoro}, Campos, Delchambre, Dell'Oro, Esquej, {Fern{\'a}ndez-Hern{\'a}ndez}, Fraile, Garabato, {Garc{\'i}a-Lario}, Gosset, Haigron, Halbwachs, Hambly, Harrison, Hern{\'a}ndez, Hestroffer, Hodgkin, Holl, Jan{\ss}en, {de
  Fombelle}, Jordan, {Krone-Martins}, Lanzafame, L{\"o}ffler, Marchal, Marrese, Moitinho, Muinonen, Osborne, Pancino, Pauwels, {Recio-Blanco}, Reyl{\'e}, Riello, Rimoldini, Roegiers, Rybizki, Sarro, Siopis, Smith, Sozzetti, Utrilla, {van Leeuwen}, Abbas, {\'A}brah{\'a}m, Aramburu, Aerts, Aguado, Ajaj, {Aldea-Montero}, Altavilla, {\'A}lvarez, Alves, Anders, Anderson, Varela, Antoja, Baines, Baker, {Balaguer-N{\'u}{\~n}ez}, Balbinot, Balog, Barache, Barbato, Barros, Barstow, Bartolom{\'e}, Bassilana, Bauchet, Becciani, Bellazzini, Berihuete, Bernet, Bertone, Bianchi, Binnenfeld, {Blanco-Cuaresma}, Blazere, Boch, Bombrun, Bossini, Bouquillon, Bragaglia, Bramante, Breedt, Bressan, Brouillet, Brugaletta, Bucciarelli, Burlacu, Butkevich, Buzzi, Caffau, Cancelliere, {Cantat-Gaudin}, Carballo, Carlucci, Carnerero, Carrasco, Casamiquela, Castellani, {Castro-Ginard}, Chaoul, Charlot, Chemin, Chiaramida, Chiavassa, Chornay, Comoretto, Contursi, Cooper, Cornez, Cowell, Crifo, Cropper, Crosta, Crowley, Dafonte,
  Dapergolas, David, David, {de Laverny}, De~Luise, De~March, De~Ridder, {de Souza}, {de Torres}, {del Peloso}, {del Pozo}, Delbo, Delgado, Delisle, Demouchy, Dharmawardena, Di~Matteo, Diakite, Diener, Distefano, Dolding, Edvardsson, Enke, Fabre, Fabrizio, Faigler, Fedorets, Fernique, Fienga, Figueras, Fournier, Fouron, Fragkoudi, Gai, {Garcia-Gutierrez}, {Garcia-Reinaldos}, {Garc{\'i}a-Torres}, Garofalo, Gavel, Gavras, Gerlach, Geyer, Giacobbe, Gilmore, Girona, Giuffrida, Gomel, Gomez, {Gonz{\'a}lez-N{\'u}{\~n}ez}, {Gonz{\'a}lez-Santamar{\'i}a}, {Gonz{\'a}lez-Vidal}, Granvik, Guillout, Guiraud, {Guti{\'e}rrez-S{\'a}nchez}, Guy, Hatzidimitriou, Hauser, Haywood, Helmer, Helmi, Sarmiento, Hidalgo, Hilger, H{\l}adczuk, Hobbs, Holland, Huckle, Jardine, Jasniewicz, Piccolo, {Jim{\'e}nez-Arranz}, Jorissen, Campillo, Julbe, Karbevska, Kervella, Khanna, Kontizas, Kordopatis, Korn, K{\'o}sp{\'a}l, {Kostrzewa-Rutkowska}, Kruszy{\'n}ska, Kun, Laizeau, Lambert, Lanza, Lasne, Campion, Lebreton, Lebzelter, Leccia, Leclerc,
  {Lecoeur-Taibi}, Liao, Licata, Lindstr{\o}m, Lister, Livanou, Lobel, Lorca, Loup, Pardo, Romeo, Managau, Mann, Manteiga, Marchant, Marconi, Marcos, Santos, Pina, Marinoni, Marocco, Marshall, Polo, {Mart{\'i}n-Fleitas}, Marton, Mary, Masip, Massari, {Mastrobuono-Battisti}, Mazeh, McMillan, Messina, Michalik, Millar, Mints, Molina, Molinaro, Moln{\'a}r, Monari, Mongui{\'o}, Montegriffo, Montero, Mor, Mora, Morbidelli, Morel, Morris, Muraveva, Murphy, Musella, Nagy, Noval, Oca{\~n}a, Ogden, Ordenovic, Osinde, Pagani, Pagano, Palaversa, Palicio, {Pallas-Quintela}, Panahi, {Payne-Wardenaar}, Esteller, Penttil{\"a}, Pichon, Piersimoni, Pineau, Plachy, Plum, Poggio, Pr{\v s}a, Pulone, Racero, Ragaini, Rainer, Raiteri, Rambaux, Ramos, {Ramos-Lerate}, Fiorentin, Regibo, Richards, Diaz, Ripepi, Riva, Rix, Rixon, Robichon, Robin, Robin, Roelens, Rogues, Rohrbasser, {Romero-G{\'o}mez}, Rowell, Royer, Mieres, Rybicki, Sadowski, N{\'u}{\~n}ez, Sell{\'e}s, Sahlmann, Salguero, Samaras, Gimenez, Sanna, Santove{\~n}a,
  Sarasso, Schultheis, Sciacca, Segol, Segovia, S{\'e}gransan, Semeux, Shahaf, Siddiqui, Siebert, Siltala, Silvelo, Slezak, Slezak, Smart, Snaith, Solano, Solitro, Souami, Souchay, Spagna, Spina, Spoto, Steele, Steidelm{\"u}ller, Stephenson, S{\"u}veges, Surdej, Szabados, {Szegedi-Elek}, Taris, Taylo, Teixeira, Tolomei, Tonello, Torra, Torra, Elipe, Trabucchi, Tsounis, Turon, Ulla, Unger, Vaillant, {van Dillen}, {van Reeven}, Vanel, Vecchiato, Viala, Vicente, Voutsinas, Weiler, Wevers, Wyrzykowski, Yoldas, Yvard, Zhao, Zorec, Zucker, \& Zwitter}]{gaiacollaborationGaiaDataRelease2022}
Collaboration, G., Vallenari, A., Brown, A. G.~A., {et~al.} 2022{\natexlab{a}}, Gaia {{Data Release}} 3: {{Summary}} of the Content and Survey Properties, \dodoi{10.1051/0004-6361/202243940}

\bibitem[{Collaboration {et~al.}(2018)Collaboration, {Price-Whelan}, Sip{\H o}cz, G{\"u}nther, Lim, Crawford, Conseil, Shupe, Craig, Dencheva, Ginsburg, VanderPlas, Bradley, {P{\'e}rez-Su{\'a}rez}, {de Val-Borro}, Aldcroft, Cruz, Robitaille, Tollerud, Ardelean, Babej, Bachetti, Bakanov, Bamford, Barentsen, Barmby, Baumbach, Berry, Biscani, Boquien, Bostroem, Bouma, Brammer, Bray, Breytenbach, Buddelmeijer, Burke, Calderone, Rodr{\'i}guez, Cara, Cardoso, Cheedella, Copin, Crichton, D{\'A}vella, Deil, Depagne, Dietrich, Donath, Droettboom, Earl, Erben, Fabbro, Ferreira, Finethy, Fox, Garrison, Gibbons, Goldstein, Gommers, Greco, Greenfield, Groener, Grollier, Hagen, Hirst, Homeier, Horton, Hosseinzadeh, Hu, Hunkeler, Ivezi{\'c}, Jain, Jenness, Kanarek, Kendrew, Kern, Kerzendorf, Khvalko, King, Kirkby, Kulkarni, Kumar, Lee, Lenz, Littlefair, Ma, Macleod, Mastropietro, McCully, Montagnac, Morris, Mueller, Mumford, Muna, Murphy, Nelson, Nguyen, Ninan, N{\"o}the, Ogaz, Oh, Parejko, Parley, Pascual, Patil, Patil,
  Plunkett, Prochaska, Rastogi, Janga, Sabater, Sakurikar, Seifert, Sherbert, {Sherwood-Taylor}, Shih, Sick, Silbiger, Singanamalla, Singer, Sladen, Sooley, Sornarajah, Streicher, Teuben, Thomas, Tremblay, Turner, Terr{\'o}n, {van Kerkwijk}, {de la Vega}, Watkins, Weaver, Whitmore, Woillez, \& Zabalza}]{theastropycollaborationAstropyProjectBuilding2018}
Collaboration, T.~A., {Price-Whelan}, A.~M., Sip{\H o}cz, B.~M., {et~al.} 2018, AJ, 156, 123, \dodoi{10.3847/1538-3881/aabc4f}

\bibitem[{Collaboration {et~al.}(2022{\natexlab{b}})Collaboration, {Price-Whelan}, Lim, Earl, Starkman, Bradley, Shupe, Patil, Corrales, Brasseur, N{\"o}the, Donath, Tollerud, Morris, Ginsburg, Vaher, Weaver, Tocknell, Jamieson, {van Kerkwijk}, Robitaille, Merry, Bachetti, G{\"u}nther, Aldcroft, {Alvarado-Montes}, Archibald, B{\'o}di, Bapat, Barentsen, Baz{\'a}n, Biswas, Boquien, Burke, Cara, Cara, Conroy, Conseil, Craig, Cross, Cruz, D'Eugenio, Dencheva, Devillepoix, Dietrich, Eigenbrot, Erben, Ferreira, {Foreman-Mackey}, Fox, Freij, Garg, Geda, Glattly, Gondhalekar, Gordon, Grant, Greenfield, Groener, Guest, Gurovich, Handberg, Hart, {Hatfield-Dodds}, Homeier, Hosseinzadeh, Jenness, Jones, Joseph, Kalmbach, Karamehmetoglu, Ka{\l}uszy{\'n}ski, Kelley, Kern, Kerzendorf, Koch, Kulumani, Lee, Ly, Ma, MacBride, Maljaars, Muna, Murphy, Norman, O'Steen, Oman, Pacifici, Pascual, {Pascual-Granado}, Patil, Perren, Pickering, Rastogi, Roulston, Ryan, Rykoff, Sabater, Sakurikar, Salgado, Sanghi, Saunders, Savchenko,
  Schwardt, {Seifert-Eckert}, Shih, Jain, Shukla, Sick, Simpson, Singanamalla, Singer, Singhal, Sinha, Sip{\H o}cz, Spitler, Stansby, Streicher, {\v S}umak, Swinbank, Taranu, Tewary, Tremblay, {de Val-Borro}, Van~Kooten, Vasovi{\'c}, Verma, Cardoso, Williams, Wilson, Winkel, {Wood-Vasey}, Xue, Yoachim, ZHANG, \& Zonca}]{theastropycollaborationAstropyProjectSustaining2022}
Collaboration, T.~A., {Price-Whelan}, A.~M., Lim, P.~L., {et~al.} 2022{\natexlab{b}}, The {{Astropy Project}}: {{Sustaining}} and {{Growing}} a {{Community-oriented Open-source Project}} and the {{Latest Major Release}} (v5.0) of the {{Core Package}}, \dodoi{10.3847/1538-4357/ac7c74}

\bibitem[{Costantin {et~al.}(2023)Costantin, {P{\'e}rez-Gonz{\'a}lez}, Guo, Buttitta, Jogee, Bagley, Barro, Kartaltepe, Koekemoer, Cabello, Corsini, {M{\'e}ndez-Abreu}, {de la Vega}, Iyer, Bisigello, Cheng, Morelli, Arrabal~Haro, Buitrago, Cooper, Dekel, Dickinson, Finkelstein, Giavalisco, Holwerda, {Huertas-Company}, Lucas, Papovich, Pirzkal, Seill{\'e}, {Vega-Ferrero}, Wuyts, \& Yung}]{costantinMilkyWaylikeBarred2023}
Costantin, L., {P{\'e}rez-Gonz{\'a}lez}, P.~G., Guo, Y., {et~al.} 2023, Nature, 623, 499, \dodoi{10.1038/s41586-023-06636-x}

\bibitem[{Debattista {et~al.}(2017)Debattista, Ness, Gonzalez, Freeman, Zoccali, \& Minniti}]{debattistaSeparationStellarPopulations2017}
Debattista, V.~P., Ness, M., Gonzalez, O.~A., {et~al.} 2017, Monthly Notices of the Royal Astronomical Society, 469, 1587, \dodoi{10.1093/mnras/stx947}

\bibitem[{D{\'e}k{\'a}ny {et~al.}(2013)D{\'e}k{\'a}ny, Minniti, Catelan, Zoccali, Saito, Hempel, \& Gonzalez}]{dekanyVVVSurveyNearinfrared2013}
D{\'e}k{\'a}ny, I., Minniti, D., Catelan, M., {et~al.} 2013, The Astrophysical Journal, 776, L19, \dodoi{10.1088/2041-8205/776/2/L19}

\bibitem[{Di~Matteo(2016)}]{dimatteoDiscOriginMilky2016a}
Di~Matteo, P. 2016, Publ. Astron. Soc. Aust., 33, e027, \dodoi{10.1017/pasa.2016.11}

\bibitem[{Di~Matteo {et~al.}(2019)Di~Matteo, Haywood, Lehnert, Katz, Khoperskov, Snaith, G{\'o}mez, \& Robichon}]{dimatteoMilkyWayHas2019}
Di~Matteo, P., Haywood, M., Lehnert, M.~D., {et~al.} 2019, Astronomy and Astrophysics, 632, A4, \dodoi{10.1051/0004-6361/201834929}

\bibitem[{Drimmel {et~al.}(2023)Drimmel, {Romero-G{\'o}mez}, Chemin, Ramos, Poggio, Ripepi, Andrae, Blomme, {Cantat-Gaudin}, {Castro-Ginard}, Clementini, Figueras, Fouesneau, Fr{\'e}mat, Jardine, Khanna, Lobel, Marshall, Muraveva, Brown, Vallenari, Prusti, de~Bruijne, Arenou, Babusiaux, Biermann, Creevey, Ducourant, Evans, Eyer, Guerra, Hutton, Jordi, Klioner, Lammers, Lindegren, Luri, Mignard, Panem, Pourbaix, Randich, Sartoretti, Soubiran, Tanga, Walton, {Bailer-Jones}, Bastian, Jansen, Katz, Lattanzi, van Leeuwen, Bakker, Cacciari, Casta{\~n}eda, Angeli, Fabricius, Galluccio, Guerrier, Heiter, Masana, Messineo, Mowlavi, Nicolas, Nienartowicz, Pailler, Panuzzo, Riclet, Roux, Seabroke, Sordo, Th{\'e}venin, {Gracia-Abril}, Portell, Teyssier, Altmann, Audard, {Bellas-Velidis}, Benson, Berthier, Burgess, Busonero, Busso, C{\'a}novas, Carry, Cellino, Cheek, Damerdji, Davidson, de~Teodoro, Campos, Delchambre, Dell'Oro, Esquej, {Fern{\'a}ndez-Hern{\'a}ndez}, Fraile, Garabato, {Garc{\'i}a-Lario}, Gosset, Haigron,
  Halbwachs, Hambly, Harrison, Hern{\'a}ndez, Hestroffer, Hodgkin, Holl, Jan{\ss}en, de~Fombelle, Jordan, {Krone-Martins}, Lanzafame, L{\"o}ffler, Marchal, Marrese, Moitinho, Muinonen, Osborne, Pancino, Pauwels, {Recio-Blanco}, Reyl{\'e}, Riello, Rimoldini, Roegiers, Rybizki, Sarro, Siopis, Smith, Sozzetti, Utrilla, van Leeuwen, Abbas, {\'A}brah{\'a}m, Aramburu, Aerts, Aguado, Ajaj, {Aldea-Montero}, Altavilla, {\'A}lvarez, Alves, Anders, Anderson, Varela, Antoja, Baines, Baker, {Balaguer-N{\'u}{\~n}ez}, Balbinot, Balog, Barache, Barbato, Barros, Barstow, Bartolom{\'e}, Bassilana, Bauchet, Becciani, Bellazzini, Berihuete, Bernet, Bertone, Bianchi, Binnenfeld, {Blanco-Cuaresma}, Boch, Bombrun, Bossini, Bouquillon, Bragaglia, Bramante, Breedt, Bressan, Brouillet, Brugaletta, Bucciarelli, Burlacu, Butkevich, Buzzi, Caffau, Cancelliere, Carballo, Carlucci, Carnerero, Carrasco, Casamiquela, Castellani, Chaoul, Charlot, Chiaramida, Chiavassa, Chornay, Comoretto, Contursi, Cooper, Cornez, Cowell, Crifo, Cropper,
  Crosta, Crowley, Dafonte, Dapergolas, David, de~Laverny, Luise, March, Ridder, de~Souza, de~Torres, del Peloso, del Pozo, Delbo, Delgado, Delisle, Demouchy, Dharmawardena, Matteo, Diakite, Diener, Distefano, Dolding, Enke, Fabre, Fabrizio, Faigler, Fedorets, Fernique, Fournier, Fouron, Fragkoudi, Gai, {Garcia-Gutierrez}, {Garcia-Reinaldos}, {Garc{\'i}a-Torres}, Garofalo, Gavel, Gavras, Gerlach, Geyer, Giacobbe, Gilmore, Girona, Giuffrida, Gomel, Gomez, {Gonz{\'a}lez-N{\'u}{\~n}ez}, {Gonz{\'a}lez-Santamar{\'i}a}, {Gonz{\'a}lez-Vidal}, Granvik, Guillout, Guiraud, {Guti{\'e}rrez-S{\'a}nchez}, Guy, Hatzidimitriou, Hauser, Haywood, Helmer, Helmi, Sarmiento, Hidalgo, H{\l}adczuk, Hobbs, Holland, Huckle, Jasniewicz, Piccolo, {Jim{\'e}nez-Arranz}, Campillo, Julbe, Karbevska, Kervella, Kordopatis, Korn, K{\'o}sp{\'a}l, {Kostrzewa-Rutkowska}, Kruszy{\'n}ska, Kun, Laizeau, Lambert, Lanza, Lasne, Campion, Lebreton, Lebzelter, Leccia, Leclerc, {Lecoeur-Taibi}, Liao, Licata, Lindstr{\o}m, Lister, Livanou, Lorca, Loup,
  Pardo, Romeo, Managau, Mann, Manteiga, Marchant, Marconi, Marcos, Santos, Pina, Marinoni, Marocco, Polo, {Mart{\'i}n-Fleitas}, Marton, Mary, Masip, Massari, {Mastrobuono-Battisti}, Mazeh, McMillan, Messina, Michalik, Millar, Mints, Molina, Molinaro, Moln{\'a}r, Monari, Mongui{\'o}, Montegriffo, Montero, Mor, Mora, Morbidelli, Morel, Morris, Murphy, Musella, Nagy, Noval, Oca{\~n}a, Ogden, Ordenovic, Osinde, Pagani, Pagano, Palaversa, Palicio, {Pallas-Quintela}, Panahi, {Payne-Wardenaar}, Esteller, Penttil{\"a}, Pichon, Piersimoni, Pineau, Plachy, Plum, Pr{\v s}a, Pulone, Racero, Ragaini, Rainer, Raiteri, {Ramos-Lerate}, Fiorentin, Regibo, Richards, Diaz, Riva, Rix, Rixon, Robichon, Robin, Robin, Roelens, Rogues, Rohrbasser, Rowell, Royer, Mieres, Rybicki, Sadowski, N{\'u}{\~n}ez, Sell{\'e}s, Sahlmann, Salguero, Samaras, Gimenez, Sanna, Santove{\~n}a, Sarasso, Schultheis, Sciacca, Segol, Segovia, S{\'e}gransan, Semeux, Shahaf, Siddiqui, Siebert, Siltala, Silvelo, Slezak, Slezak, Smart, Snaith, Solano,
  Solitro, Souami, Souchay, Spagna, Spina, Spoto, Steele, Steidelm{\"u}ller, Stephenson, S{\"u}veges, Surdej, Szabados, {Szegedi-Elek}, Taris, Taylor, Teixeira, Tolomei, Tonello, Torra, Torra, Elipe, Trabucchi, Tsounis, Turon, Ulla, Unger, Vaillant, van Dillen, van Reeven, Vanel, Vecchiato, Viala, Vicente, Voutsinas, Weiler, Wevers, Wyrzykowski, Yoldas, Yvard, Zhao, Zorec, Zucker, \& Zwitter}]{drimmelGaiaDataRelease2023}
Drimmel, R., {Romero-G{\'o}mez}, M., Chemin, L., {et~al.} 2023, A\&A, 674, A37, \dodoi{10.1051/0004-6361/202243797}

\bibitem[{Dwek {et~al.}(1995)Dwek, Arendt, Hauser, Kelsall, Lisse, Moseley, Silverberg, Sodroski, \& Weiland}]{dwekMorphologyNearinfraredLuminosity1995}
Dwek, E., Arendt, R.~G., Hauser, M.~G., {et~al.} 1995, ApJ, 445, 716, \dodoi{10.1086/175734}

\bibitem[{Eggen {et~al.}(1962)Eggen, {Lynden-Bell}, \& Sandage}]{eggenEvidenceMotionsOld1962}
Eggen, O.~J., {Lynden-Bell}, D., \& Sandage, A.~R. 1962, The Astrophysical Journal, 136, 748, \dodoi{10.1086/147433}

\bibitem[{Ferreira {et~al.}(2023)Ferreira, Conselice, Sazonova, Ferrari, Caruana, Tohill, Lucatelli, Adams, Irodotou, Marshall, Roper, Lovell, Verma, Austin, Trussler, \& Wilkins}]{ferreiraJWSTHubbleSequence2023}
Ferreira, L., Conselice, C.~J., Sazonova, E., {et~al.} 2023, The Astrophysical Journal, 955, 94, \dodoi{10.3847/1538-4357/acec76}

\bibitem[{Fragkoudi {et~al.}(2017)Fragkoudi, Di~Matteo, Haywood, G{\'o}mez, Combes, Katz, \& Semelin}]{fragkoudiBarsBoxyPeanut2017a}
Fragkoudi, F., Di~Matteo, P., Haywood, M., {et~al.} 2017, A\&A, 606, A47, \dodoi{10.1051/0004-6361/201630244}

\bibitem[{Fragkoudi {et~al.}(2018)Fragkoudi, Di~Matteo, Haywood, Schultheis, Khoperskov, G{\'o}mez, \& Combes}]{fragkoudiDiscOriginMilky2018}
---. 2018, Astronomy and Astrophysics, 616, A180, \dodoi{10.1051/0004-6361/201732509}

\bibitem[{Fragkoudi {et~al.}(2020)Fragkoudi, Grand, Pakmor, {Bl{\'a}zquez-Calero}, Gargiulo, Gomez, Marinacci, Monachesi, Ness, Perez, Tissera, \& White}]{fragkoudiChemodynamicsBarredGalaxies2020a}
Fragkoudi, F., Grand, R. J.~J., Pakmor, R., {et~al.} 2020, Monthly Notices of the Royal Astronomical Society, 494, 5936, \dodoi{10.1093/mnras/staa1104}

\bibitem[{Frankel {et~al.}(2019)Frankel, Sanders, Rix, Ting, \& Ness}]{frankelInsideoutGrowthGalactic2019}
Frankel, N., Sanders, J., Rix, H.-W., Ting, Y.-S., \& Ness, M. 2019, ApJ, 884, 99, \dodoi{10.3847/1538-4357/ab4254}

\bibitem[{Freeman {et~al.}(2013)Freeman, Ness, {Wylie-de-Boer}, Athanassoula, {Bland-Hawthorn}, Asplund, Lewis, Yong, Lane, Kiss, \& Ibata}]{freemanARGOSIIGalactic2013}
Freeman, K., Ness, M., {Wylie-de-Boer}, E., {et~al.} 2013, Monthly Notices of the Royal Astronomical Society, 428, 3660, \dodoi{10.1093/mnras/sts305}

\bibitem[{Fulbright {et~al.}(2006)Fulbright, McWilliam, \& Rich}]{fulbrightAbundancesBaadeWindow2006}
Fulbright, {\relax Jon}.~P., McWilliam, A., \& Rich, R.~M. 2006, The Astrophysical Journal, 636, 821, \dodoi{10.1086/498205}

\bibitem[{{Gaia Collaboration} {et~al.}(2018){Gaia Collaboration}, Katz, Antoja, {Romero-G{\'o}mez}, Drimmel, Reyl{\'e}, Seabroke, Soubiran, Babusiaux, Di~Matteo, Figueras, Poggio, Robin, Evans, Brown, Vallenari, Prusti, {de Bruijne}, {Bailer-Jones}, Biermann, Eyer, Jansen, Jordi, Klioner, Lammers, Lindegren, Luri, Mignard, Panem, Pourbaix, Randich, Sartoretti, Siddiqui, {van Leeuwen}, Walton, Arenou, Bastian, Cropper, Lattanzi, Bakker, Cacciari, {Casta n}, Chaoul, Cheek, De~Angeli, Fabricius, Guerra, Holl, Masana, Messineo, Mowlavi, Nienartowicz, Panuzzo, Portell, Riello, Tanga, Th{\'e}venin, {Gracia-Abril}, Comoretto, {Garcia-Reinaldos}, Teyssier, Altmann, Andrae, Audard, {Bellas-Velidis}, Benson, Berthier, Blomme, Burgess, Busso, Carry, Cellino, Clementini, Clotet, Creevey, Davidson, De~Ridder, Delchambre, Dell'Oro, Ducourant, {Fern{\'a}ndez-Hern{\'a}ndez}, Fouesneau, Fr{\'e}mat, Galluccio, {Garc{\'i}a-Torres}, {Gonz{\'a}lez-N{\'u}{\~n}ez}, {Gonz{\'a}lez-Vidal}, Gosset, Guy, Halbwachs, Hambly, Harrison,
  Hern{\'a}ndez, Hestroffer, Hodgkin, Hutton, Jasniewicz, {Jean-Antoine-Piccolo}, Jordan, Korn, {Krone-Martins}, Lanzafame, Lebzelter, L{\"o}ffler, Manteiga, Marrese, {Mart{\'i}n-Fleitas}, Moitinho, Mora, Muinonen, Osinde, Pancino, Pauwels, Petit, {Recio-Blanco}, Richards, Rimoldini, Sarro, Siopis, Smith, Sozzetti, S{\"u}veges, Torra, {van Reeven}, Abbas, Abreu~Aramburu, Accart, Aerts, Altavilla, {\'A}lvarez, Alvarez, Alves, Anderson, Andrei, Anglada~Varela, Antiche, Arcay, Astraatmadja, Bach, Baker, {Balaguer-N{\'u}{\~n}ez}, Balm, Barache, Barata, Barbato, Barblan, Barklem, Barrado, Barros, Barstow, Bartholom{\'e}~Mu{\~n}oz, Bassilana, Becciani, Bellazzini, Berihuete, Bertone, Bianchi, Bienaym{\'e}, {Blanco-Cuaresma}, Boch, Boeche, Bombrun, Borrachero, Bossini, Bouquillon, Bourda, Bragaglia, Bramante, Breddels, Bressan, Brouillet, Br{\"u}semeister, Brugaletta, Bucciarelli, Burlacu, Busonero, Butkevich, Buzzi, Caffau, Cancelliere, Cannizzaro, {Cantat-Gaudin}, Carballo, Carlucci, Carrasco, Casamiquela,
  Castellani, {Castro-Ginard}, Charlot, Chemin, Chiavassa, Cocozza, Costigan, Cowell, Crifo, Crosta, Crowley, Cuypers, Dafonte, Damerdji, Dapergolas, David, David, {de Laverny}, De~Luise, De~March, {de Souza}, {de Torres}, Debosscher, {del Pozo}, Delbo, Delgado, Delgado, Diakite, Diener, Distefano, Dolding, Drazinos, Dur{\'a}n, Edvardsson, Enke, Eriksson, Esquej, Eynard~Bontemps, Fabre, Fabrizio, Faigler, {Falc a}, Farr{\`a}s~Casas, Federici, Fedorets, Fernique, Filippi, Findeisen, Fonti, Fraile, Fraser, Fr{\'e}zouls, Gai, Galleti, Garabato, {Garc{\'i}a-Sedano}, Garofalo, Garralda, Gavel, Gavras, Gerssen, Geyer, Giacobbe, Gilmore, Girona, Giuffrida, Glass, Gomes, Granvik, Gueguen, Guerrier, Guiraud, Guti{\'e}, Haigron, Hatzidimitriou, Hauser, Haywood, Heiter, Helmi, Heu, Hilger, Hobbs, Hofmann, Holland, Huckle, Hypki, Icardi, Jan{\ss}en, {Jevardat de Fombelle}, Jonker, Juh{\'a}sz, Julbe, Karampelas, Kewley, Klar, Kochoska, Kohley, Kolenberg, Kontizas, Kontizas, Koposov, Kordopatis, {Kostrzewa-Rutkowska},
  Koubsky, Lambert, Lanza, Lasne, Lavigne, Le~Fustec, {Le Poncin-Lafitte}, Lebreton, Leccia, Leclerc, {Lecoeur-Taibi}, Lenhardt, Leroux, Liao, Licata, Lindstr{\o}m, Lister, Livanou, Lobel, L{\'o}pez, Managau, Mann, Mantelet, Marchal, Marchant, Marconi, Marinoni, Marschalk{\'o}, Marshall, Martino, Marton, Mary, Massari, Matijevi{\v c}, Mazeh, McMillan, Messina, Michalik, Millar, Molina, Molinaro, Moln{\'a}r, Montegriffo, Mor, Morbidelli, Morel, Morris, Mulone, Muraveva, Musella, Nelemans, Nicastro, Noval, O'Mullane, Ord{\'e}novic, {Ord{\'o}{\~n}ez-Blanco}, Osborne, Pagani, Pagano, Pailler, Palacin, Palaversa, Panahi, Pawlak, Piersimoni, Pineau, Plachy, Plum, Poujoulet, Pr{\v s}a, Pulone, Racero, Ragaini, Rambaux, {Ramos-Lerate}, Regibo, Riclet, Ripepi, Riva, Rivard, Rixon, Roegiers, Roelens, Rowell, Royer, {Ruiz-Dern}, Sadowski, Sagrist{\`a}~Sell{\'e}s, Sahlmann, Salgado, Salguero, Sanna, {Santana-Ros}, Sarasso, Savietto, Schultheis, Sciacca, Segol, Segovia, S{\'e}gransan, Shih, Siltala, Silva, Smart, Smith,
  Solano, Solitro, Sordo, Soria~Nieto, Souchay, Spagna, Spoto, Stampa, Steele, Steidelm{\"u}ller, Stephenson, Stoev, Suess, Surdej, Szabados, {Szegedi-Elek}, Tapiador, Taris, Tauran, Taylor, Teixeira, Terrett, Teyssandier, Thuillot, Titarenko, Torra~Clotet, Turon, Ulla, Utrilla, Uzzi, Vaillant, Valentini, Valette, {van Elteren}, Van~Hemelryck, {van Leeuwen}, Vaschetto, Vecchiato, Veljanoski, Viala, Vicente, Vogt, {von Essen}, Voss, Votruba, Voutsinas, Walmsley, Weiler, Wertz, Wevers, Wyrzykowski, Yoldas, {\v Z}erjal, Ziaeepour, Zorec, Zschocke, Zucker, Zurbach, \& Zwitter}]{gaiacollaborationGaiaDataRelease2018}
{Gaia Collaboration}, Katz, D., Antoja, T., {et~al.} 2018, Astronomy and Astrophysics, 616, A11, \dodoi{10.1051/0004-6361/201832865}

\bibitem[{{Gaia Collaboration} {et~al.}(2021){Gaia Collaboration}, Antoja, McMillan, Kordopatis, Ramos, Helmi, Balbinot, {Cantat-Gaudin}, Chemin, Figueras, Jordi, Khanna, {Romero-G{\'o}mez}, Seabroke, Brown, Vallenari, Prusti, {de Bruijne}, Babusiaux, Biermann, Creevey, Evans, Eyer, Hutton, Jansen, Klioner, Lammers, Lindegren, Luri, Mignard, Panem, Pourbaix, Randich, Sartoretti, Soubiran, Walton, Arenou, {Bailer-Jones}, Bastian, Cropper, Drimmel, Katz, Lattanzi, {van Leeuwen}, Bakker, Casta{\~n}eda, De~Angeli, Ducourant, Fabricius, Fouesneau, Fr{\'e}mat, Guerra, Guerrier, Guiraud, {Jean-Antoine Piccolo}, Masana, Messineo, Mowlavi, Nicolas, Nienartowicz, Pailler, Panuzzo, Riclet, Roux, Sordo, Tanga, Th{\'e}venin, {Gracia-Abril}, Portell, Teyssier, Altmann, Andrae, {Bellas-Velidis}, Benson, Berthier, Blomme, Brugaletta, Burgess, Busso, Carry, Cellino, Cheek, Clementini, Damerdji, Davidson, Delchambre, Dell'Oro, {Fern{\'a}ndez-Hern{\'a}ndez}, Galluccio, {Garc{\'i}a-Lario}, {Garcia-Reinaldos},
  {Gonz{\'a}lez-N{\'u}{\~n}ez}, Gosset, Haigron, Halbwachs, Hambly, Harrison, Hatzidimitriou, Heiter, Hern{\'a}ndez, Hestroffer, Hodgkin, Holl, Jan{\ss}en, {Jevardat de Fombelle}, Jordan, {Krone-Martins}, Lanzafame, L{\"o}ffler, Lorca, Manteiga, Marchal, Marrese, Moitinho, Mora, Muinonen, Osborne, Pancino, Pauwels, {Recio-Blanco}, Richards, Riello, Rimoldini, Robin, Roegiers, Rybizki, Sarro, Siopis, Smith, Sozzetti, Ulla, Utrilla, {van Leeuwen}, {van Reeven}, Abbas, Abreu~Aramburu, Accart, Aerts, Aguado, Ajaj, Altavilla, {\'A}lvarez, {\'A}lvarez Cid-Fuentes, Alves, Anderson, Varela, Audard, Baines, Baker, {Balaguer-N{\'u}{\~n}ez}, Balog, Barache, Barbato, Barros, Barstow, Bartolom{\'e}, Bassilana, Bauchet, {Baudesson-Stella}, Becciani, Bellazzini, Bernet, Bertone, Bianchi, {Blanco-Cuaresma}, Boch, Bombrun, Bossini, Bouquillon, Bragaglia, Bramante, Breedt, Bressan, Brouillet, Bucciarelli, Burlacu, Busonero, Butkevich, Buzzi, Caffau, Cancelliere, C{\'a}novas, Carballo, Carlucci, Carnerero, Carrasco,
  Casamiquela, Castellani, {Castro-Ginard}, Castro~Sampol, Chaoul, Charlot, Chiavassa, Cioni, Comoretto, Cooper, Cornez, Cowell, Crifo, Crosta, Crowley, Dafonte, Dapergolas, David, David, {de Laverny}, De~Luise, De~March, De~Ridder, {de Souza}, {de Teodoro}, {de Torres}, {del Peloso}, {del Pozo}, Delgado, Delgado, Delisle, Di~Matteo, Diakite, Diener, Distefano, Dolding, Eappachen, Enke, Esquej, Fabre, Fabrizio, Faigler, Fedorets, Fernique, Fienga, Fouron, Fragkoudi, Fraile, Franke, Gai, Garabato, {Garcia-Gutierrez}, {Garc{\'i}a-Torres}, Garofalo, Gavras, Gerlach, Geyer, Giacobbe, Gilmore, Girona, Giuffrida, Gomez, {Gonzalez-Santamaria}, {Gonz{\'a}lez-Vidal}, Granvik, {Guti{\'e}rrez-S{\'a}nchez}, Guy, Hauser, Haywood, Hidalgo, Hilger, H{\l}adczuk, Hobbs, Holland, Huckle, Jasniewicz, Jonker, Juaristi~Campillo, Julbe, Karbevska, Kervella, Kochoska, Kontizas, Korn, {Kostrzewa-Rutkowska}, Kruszy{\'n}ska, Lambert, Lanza, Lasne, Le~Campion, Le~Fustec, Lebreton, Lebzelter, Leccia, Leclerc, {Lecoeur-Taibi}, Liao,
  Licata, Lindstr{\o}m, Lister, Livanou, Lobel, Madrero~Pardo, Managau, Mann, Marchant, Marconi, Marcos~Santos, Marinoni, Marocco, Marshall, Martin~Polo, {Mart{\'i}n-Fleitas}, Masip, Massari, {Mastrobuono-Battisti}, Mazeh, Messina, Michalik, Millar, Mints, Molina, Molinaro, Moln{\'a}r, Montegriffo, Mor, Morbidelli, Morel, Morris, Mulone, Munoz, Muraveva, Murphy, Musella, Noval, Ord{\'e}novic, Orr{\`u}, Osinde, Pagani, Pagano, Palaversa, Palicio, Panahi, Pawlak, Pe{\~n}alosa~Esteller, Penttil{\"a}, Piersimoni, Pineau, Plachy, Plum, Poggio, Poretti, Poujoulet, Pr{\v s}a, Pulone, Racero, Ragaini, Rainer, Raiteri, Rambaux, {Ramos-Lerate}, Re~Fiorentin, Regibo, Reyl{\'e}, Ripepi, Riva, Rixon, Robichon, Robin, Roelens, Rohrbasser, Rowell, Royer, Rybicki, Sadowski, Sagrist{\`a}~Sell{\'e}s, Sahlmann, Salgado, Salguero, Samaras, Sanchez~Gimenez, Sanna, Santove{\~n}a, Sarasso, Schultheis, Sciacca, Segol, Segovia, S{\'e}gransan, Semeux, Siddiqui, Siebert, Siltala, Slezak, Smart, Solano, Solitro, Souami, Souchay, Spagna,
  Spoto, Steele, Steidelm{\"u}ller, Stephenson, S{\"u}veges, Szabados, {Szegedi-Elek}, Taris, Tauran, Taylor, Teixeira, Thuillot, Tonello, Torra, Torra, Turon, Unger, Vaillant, {van Dillen}, Vanel, Vecchiato, Viala, Vicente, Voutsinas, Weiler, Wevers, Wyrzykowski, Yoldas, Yvard, Zhao, Zorec, Zucker, Zurbach, \& Zwitter}]{gaiacollaborationGaiaEarlyData2021}
{Gaia Collaboration}, Antoja, T., McMillan, P.~J., {et~al.} 2021, Astronomy and Astrophysics, 649, A8, \dodoi{10.1051/0004-6361/202039714}

\bibitem[{Ghosh {et~al.}(2023)Ghosh, Fragkoudi, Di~Matteo, \& Saha}]{ghoshBarsBoxyPeanut2023}
Ghosh, S., Fragkoudi, F., Di~Matteo, P., \& Saha, K. 2023, A\&A, 674, A128, \dodoi{10.1051/0004-6361/202245275}

\bibitem[{Gonzalez {et~al.}(2015)Gonzalez, Zoccali, Vasquez, Hill, Rejkuba, Valenti, {Rojas-Arriagada}, Renzini, Babusiaux, Minniti, \& Brown}]{gonzalezGIRAFFEInnerBulge2015}
Gonzalez, O.~A., Zoccali, M., Vasquez, S., {et~al.} 2015, Astronomy and Astrophysics, 584, A46, \dodoi{10.1051/0004-6361/201526737}

\bibitem[{Grand {et~al.}(2024)Grand, Fragkoudi, G{\'o}mez, Jenkins, Marinacci, Pakmor, \& Springel}]{grandOverviewPublicData2024}
Grand, R. J.~J., Fragkoudi, F., G{\'o}mez, F.~A., {et~al.} 2024, Overview and Public Data Release of the {{Auriga Project}}: Cosmological Simulations of Dwarf and {{Milky Way-mass}} Galaxies,  arXiv.
\newblock \doeprint{2401.08750}

\bibitem[{Grand {et~al.}(2017)Grand, G{\'o}mez, Marinacci, Pakmor, Springel, Campbell, Frenk, Jenkins, \& White}]{grandAurigaProjectProperties2017}
Grand, R. J.~J., G{\'o}mez, F.~A., Marinacci, F., {et~al.} 2017, Monthly Notices of the Royal Astronomical Society, 467, 179, \dodoi{10.1093/mnras/stx071}

\bibitem[{Grand {et~al.}(2018)Grand, Bustamante, G{\'o}mez, Kawata, Marinacci, Pakmor, Rix, Simpson, Sparre, \& Springel}]{grandOriginChemicallyDistinct2018}
Grand, R. J.~J., Bustamante, S., G{\'o}mez, F.~A., {et~al.} 2018, Monthly Notices of the Royal Astronomical Society, 474, 3629, \dodoi{10.1093/mnras/stx3025}

\bibitem[{Grand {et~al.}(2020)Grand, Kawata, Belokurov, Deason, Fattahi, Fragkoudi, G{\'o}mez, Marinacci, \& Pakmor}]{grandDualOriginGalactic2020}
Grand, R. J.~J., Kawata, D., Belokurov, V., {et~al.} 2020, Monthly Notices of the Royal Astronomical Society, 497, 1603, \dodoi{10.1093/mnras/staa2057}

\bibitem[{Guo {et~al.}(2023)Guo, Jogee, Finkelstein, Chen, Wise, Bagley, Barro, Wuyts, Kocevski, Kartaltepe, McGrath, Ferguson, Mobasher, Giavalisco, Lucas, Zavala, Lotz, Grogin, {Huertas-Company}, {Vega-Ferrero}, Hathi, Arrabal~Haro, Dickinson, Koekemoer, Papovich, Pirzkal, Yung, Backhaus, Bell, Calabr{\`o}, Cleri, Coogan, Cooper, Costantin, Croton, Davis, Dekel, Franco, Gardner, Holwerda, Hutchison, Pandya, {P{\'e}rez-Gonz{\'a}lez}, Ravindranath, Rose, Trump, {de la Vega}, \& Wang}]{guoFirstLookBars2023}
Guo, Y., Jogee, S., Finkelstein, S.~L., {et~al.} 2023, The Astrophysical Journal, 945, L10, \dodoi{10.3847/2041-8213/acacfb}

\bibitem[{Hayden {et~al.}(2015)Hayden, Bovy, Holtzman, Nidever, Bird, Weinberg, Andrews, Majewski, Allende~Prieto, Anders, Beers, Bizyaev, Chiappini, Cunha, Frinchaboy, {Garc{\'i}a-Her{\'n}andez}, Garc{\'i}a~P{\'e}rez, Girardi, Harding, Hearty, Johnson, M{\'e}sz{\'a}ros, Minchev, O'Connell, Pan, Robin, Schiavon, Schneider, Schultheis, Shetrone, Skrutskie, Steinmetz, Smith, Wilson, Zamora, \& Zasowski}]{haydenChemicalCartographyAPOGEE2015}
Hayden, M.~R., Bovy, J., Holtzman, J.~A., {et~al.} 2015, The Astrophysical Journal, 808, 132, \dodoi{10.1088/0004-637X/808/2/132}

\bibitem[{Helmi {et~al.}(2018)Helmi, Babusiaux, Koppelman, Massari, Veljanoski, \& Brown}]{helmiMergerThatLed2018}
Helmi, A., Babusiaux, C., Koppelman, H.~H., {et~al.} 2018, Nature, 563, 85, \dodoi{10.1038/s41586-018-0625-x}

\bibitem[{Hill {et~al.}(2011)Hill, Lecureur, G{\'o}mez, Zoccali, Schultheis, Babusiaux, Royer, Barbuy, Arenou, Minniti, \& Ortolani}]{hillMetallicityDistributionBulge2011}
Hill, V., Lecureur, A., G{\'o}mez, A., {et~al.} 2011, A\&A, 534, A80, \dodoi{10.1051/0004-6361/200913757}

\bibitem[{Horta {et~al.}(2020)Horta, Schiavon, Mackereth, Pfeffer, Mason, Kisku, Fragkoudi, Allende~Prieto, Cunha, Hasselquist, Holtzman, Majewski, Nataf, O'Connell, Schultheis, \& Smith}]{hortaEvidenceAPOGEEPresence2020}
Horta, D., Schiavon, R.~P., Mackereth, J.~T., {et~al.} 2020, Monthly Notices of the Royal Astronomical Society, 500, 1385, \dodoi{10.1093/mnras/staa2987}

\bibitem[{Howard {et~al.}(2009)Howard, Rich, Clarkson, Mallery, Kormendy, De~Propris, Robin, Fux, Reitzel, Zhao, Kuijken, \& Koch}]{howardKINEMATICSEDGEGALACTIC2009}
Howard, C.~D., Rich, R.~M., Clarkson, W., {et~al.} 2009, ApJ, 702, L153, \dodoi{10.1088/0004-637X/702/2/L153}

\bibitem[{J{\"o}nsson {et~al.}(2020)J{\"o}nsson, Holtzman, Prieto, Cunha, {Garc{\'i}a-Hern{\'a}ndez}, Hasselquist, Masseron, Osorio, Shetrone, Smith, Stringfellow, Bizyaev, Edvardsson, Majewski, M{\'e}sz{\'a}ros, Souto, Zamora, Beaton, Bovy, Donor, Pinsonneault, Poovelil, \& Sobeck}]{jonssonAPOGEEDataSpectral2020}
J{\"o}nsson, H., Holtzman, J.~A., Prieto, C.~A., {et~al.} 2020, AJ, 160, 120, \dodoi{10.3847/1538-3881/aba592}

\bibitem[{Kormendy \& Fisher(2008)}]{kormendySecularEvolutionDisk2008}
Kormendy, J., \& Fisher, D.~B. 2008, 396, 297, \dodoi{10.48550/arXiv.0810.2534}

\bibitem[{Kormendy \& Kennicutt(2004)}]{kormendySecularEvolutionFormation2004}
Kormendy, J., \& Kennicutt, R.~C. 2004, Annu. Rev. Astron. Astrophys., 42, 603, \dodoi{10.1146/annurev.astro.42.053102.134024}

\bibitem[{Kunder {et~al.}(2012)Kunder, Koch, Rich, de~Propris, Howard, Stubbs, Johnson, Shen, Wang, Robin, Kormendy, Soto, Frinchaboy, Reitzel, Zhao, \& Origlia}]{kunderBULGERADIALVELOCITY2012}
Kunder, A., Koch, A., Rich, R.~M., {et~al.} 2012, AJ, 143, 57, \dodoi{10.1088/0004-6256/143/3/57}

\bibitem[{Le~Conte {et~al.}(2023)Le~Conte, Gadotti, Ferreira, Conselice, {de S{\'a}-Freitas}, Kim, Neumann, Fragkoudi, Athanassoula, \& Adams}]{leconteJWSTInvestigationBar2023}
Le~Conte, Z.~A., Gadotti, D.~A., Ferreira, L., {et~al.} 2023, A {{JWST}} Investigation into the Bar Fraction at Redshifts 1 {$<$} z {$<$} 3, \dodoi{10.48550/arXiv.2309.10038}

\bibitem[{Lee {et~al.}(2023)Lee, Lee, Kim, Beers, \& An}]{leeChemodynamicalAnalysisMetalrich2023}
Lee, A., Lee, Y.~S., Kim, Y.~K., Beers, T.~C., \& An, D. 2023, ApJ, 945, 56, \dodoi{10.3847/1538-4357/acb6f5}

\bibitem[{Li {et~al.}(2022)Li, Aoki, Matsuno, Xing, Suda, Tominaga, Chen, Honda, Ishigaki, Shi, Zhao, \& Zhao}]{liFourhundredVeryMetalpoor2022}
Li, H., Aoki, W., Matsuno, T., {et~al.} 2022, ApJ, 931, 147, \dodoi{10.3847/1538-4357/ac6514}

\bibitem[{Li {et~al.}(2023)Li, Shlosman, Pfenniger, \& Heller}]{liEvolutionStellarBars2023}
Li, X., Shlosman, I., Pfenniger, D., \& Heller, C. 2023, Evolution of {{Stellar Bars}} in {{Spinning Dark Matter Halos}} and {{Stellar Bulges}},  arXiv.
\newblock \doeprint{2310.01411}

\bibitem[{Lim {et~al.}(2021)Lim, {Koch-Hansen}, Chung, Johnson, Kunder, Simion, Rich, Clarkson, Pilachowski, Michael, Vivas, \& Young}]{limBlancoDECamBulge2021}
Lim, D., {Koch-Hansen}, A.~J., Chung, C., {et~al.} 2021, Astronomy and Astrophysics, 647, A34, \dodoi{10.1051/0004-6361/202039955}

\bibitem[{{L{\'o}pez-Corredoira} {et~al.}(2005){L{\'o}pez-Corredoira}, {Cabrera-Lavers}, \& Gerhard}]{lopez-corredoiraBoxyBulgeMilky2005}
{L{\'o}pez-Corredoira}, M., {Cabrera-Lavers}, A., \& Gerhard, O.~E. 2005, A\&A, 439, 107, \dodoi{10.1051/0004-6361:20053075}

\bibitem[{Lucey {et~al.}(2020)Lucey, Hawkins, Ness, Debattista, Luna, Asplund, Bensby, Casagrande, Feltzing, Freeman, Kobayashi, \& Marino}]{luceyCOMBSSurveyII2020}
Lucey, M., Hawkins, K., Ness, M., {et~al.} 2020, arXiv e-prints, 2009, arXiv:2009.03886

\bibitem[{Maihara {et~al.}(1978)Maihara, Oda, Sugiyama, \& Okuda}]{maihara4MicronObservationGalaxy1978}
Maihara, T., Oda, N., Sugiyama, T., \& Okuda, H. 1978, Publications of the Astronomical Society of Japan, 30, 1

\bibitem[{Majewski {et~al.}(2017)Majewski, Schiavon, Frinchaboy, Prieto, Barkhouser, Bizyaev, Blank, Brunner, Burton, Carrera, Chojnowski, Cunha, Epstein, Fitzgerald, P{\'e}rez, Hearty, Henderson, Holtzman, Johnson, Lam, Lawler, Maseman, M{\'e}sz{\'a}ros, Nelson, Nguyen, Nidever, Pinsonneault, Shetrone, Smee, Smith, Stolberg, Skrutskie, Walker, Wilson, Zasowski, Anders, Basu, Beland, Blanton, Bovy, Brownstein, Carlberg, Chaplin, Chiappini, Eisenstein, Elsworth, Feuillet, Fleming, {Galbraith-Frew}, Garc{\'i}a, {Garc{\'i}a-Hern{\'a}ndez}, Gillespie, Girardi, Gunn, Hasselquist, Hayden, Hekker, Ivans, Kinemuchi, Klaene, Mahadevan, Mathur, Mosser, Muna, Munn, Nichol, O'Connell, Parejko, Robin, {Rocha-Pinto}, Schultheis, Serenelli, Shane, Aguirre, Sobeck, Thompson, Troup, Weinberg, \& Zamora}]{majewskiApachePointObservatory2017}
Majewski, S.~R., Schiavon, R.~P., Frinchaboy, P.~M., {et~al.} 2017, AJ, 154, 94, \dodoi{10.3847/1538-3881/aa784d}

\bibitem[{Marchetti {et~al.}(2023)Marchetti, Joyce, Johnson, Rich, Clarkson, Kunder, Simion, \& Pilachowski}]{marchettiBlancoDECamBulge2023}
Marchetti, T., Joyce, M., Johnson, C., {et~al.} 2023, The {{Blanco DECam Bulge Survey}} ({{BDBS}}) {{VIII}}: {{Chemo-kinematics}} in the Southern {{Galactic}} Bulge from 2.3 Million Red Clump Stars with {{Gaia DR3}} Proper Motions,  arXiv.
\newblock \doeprint{2310.17542}

\bibitem[{McWilliam \& Rich(1994)}]{mcwilliamFirstDetailedAbundance1994}
McWilliam, A., \& Rich, R.~M. 1994, ApJS, 91, 749, \dodoi{10.1086/191954}

\bibitem[{Minniti {et~al.}(1995)Minniti, Olszewski, Liebert, White, Hill, \& Irwin}]{minnitiMetallicityGradientGalactic1995}
Minniti, D., Olszewski, E.~W., Liebert, J., {et~al.} 1995, Monthly Notices of the Royal Astronomical Society, 277, 1293, \dodoi{10.1093/mnras/277.4.1293}

\bibitem[{Montalb{\'a}n {et~al.}(2021)Montalb{\'a}n, Mackereth, Miglio, Vincenzo, Chiappini, Buldgen, Mosser, Noels, Scuflaire, Vrard, Willett, Davies, Hall, Bo~Nielsen, Khan, Rendle, {van Rossem}, Ferguson, \& Chaplin}]{montalbanChronologicallyDatingEarly2021}
Montalb{\'a}n, J., Mackereth, J.~T., Miglio, A., {et~al.} 2021, Nat Astron, \dodoi{10.1038/s41550-021-01347-7}

\bibitem[{Ness \& Freeman(2016)}]{nessMetallicityDistributionMilky2016}
Ness, M., \& Freeman, K. 2016, Publications of the Astronomical Society of Australia, 33, e022, \dodoi{10.1017/pasa.2015.51}

\bibitem[{Ness {et~al.}(2013{\natexlab{a}})Ness, Freeman, Athanassoula, {Wylie-de-Boer}, {Bland-Hawthorn}, Asplund, Lewis, Yong, Lane, \& Kiss}]{nessARGOSIIIStellar2013}
Ness, M., Freeman, K., Athanassoula, E., {et~al.} 2013{\natexlab{a}}, Monthly Notices of the Royal Astronomical Society, 430, 836, \dodoi{10.1093/mnras/sts629}

\bibitem[{Ness {et~al.}(2013{\natexlab{b}})Ness, Freeman, Athanassoula, {Wylie-de-Boer}, {Bland-Hawthorn}, Asplund, Lewis, Yong, Lane, Kiss, \& Ibata}]{nessARGOSIVKinematics2013}
---. 2013{\natexlab{b}}, Monthly Notices of the Royal Astronomical Society, 432, 2092, \dodoi{10.1093/mnras/stt533}

\bibitem[{Parlanti {et~al.}(2023)Parlanti, Carniani, Pallottini, Cignoni, Cresci, Kohandel, Mannucci, \& Marconi}]{parlantiALMAHintsPresence2023}
Parlanti, E., Carniani, S., Pallottini, A., {et~al.} 2023, Astronomy and Astrophysics, 673, A153, \dodoi{10.1051/0004-6361/202245603}

\bibitem[{Queiroz {et~al.}(2020)Queiroz, Anders, Chiappini, Khalatyan, Santiago, Steinmetz, Valentini, Miglio, Bossini, Barbuy, Minchev, Minniti, Garc{\'i}a~Hern{\'a}ndez, Schultheis, Beaton, Beers, Bizyaev, Brownstein, Cunha, {Fern{\'a}ndez-Trincado}, Frinchaboy, Lane, Majewski, Nataf, Nitschelm, Pan, {Roman-Lopes}, Sobeck, Stringfellow, \& Zamora}]{queirozBulgeOuterDisc2020}
Queiroz, A. B.~A., Anders, F., Chiappini, C., {et~al.} 2020, Astronomy and Astrophysics, 638, A76, \dodoi{10.1051/0004-6361/201937364}

\bibitem[{Queiroz {et~al.}(2021)Queiroz, Chiappini, {Perez-Villegas}, Khalatyan, Anders, Barbuy, Santiago, Steinmetz, Cunha, Schultheis, Majewski, Minchev, Minniti, Beaton, Cohen, {da Costa}, {Fern{\'a}ndez-Trincado}, {Garcia-Hern{\'a}ndez}, Geisler, Hasselquist, Lane, Nitschelm, {Rojas-Arriagada}, {Roman-Lopes}, Smith, \& Zasowski}]{queirozMilkyWayBar2021}
Queiroz, A. B.~A., Chiappini, C., {Perez-Villegas}, A., {et~al.} 2021, Astronomy and Astrophysics, 656, A156, \dodoi{10.1051/0004-6361/202039030}

\bibitem[{Reid {et~al.}(2014)Reid, Menten, Brunthaler, Zheng, Dame, Xu, Wu, Zhang, Sanna, Sato, Hachisuka, Choi, Immer, Moscadelli, Rygl, \& Bartkiewicz}]{reidTRIGONOMETRICPARALLAXESHIGH2014}
Reid, M.~J., Menten, K.~M., Brunthaler, A., {et~al.} 2014, ApJ, 783, 130, \dodoi{10.1088/0004-637X/783/2/130}

\bibitem[{Rich(1988)}]{richSpectroscopyAbundances881988}
Rich, R.~M. 1988, The Astronomical Journal, 95, 828, \dodoi{10.1086/114681}

\bibitem[{Rich {et~al.}(2007)Rich, Reitzel, Howard, \& Zhao}]{richBulgeRadialVelocity2007}
Rich, R.~M., Reitzel, D.~B., Howard, C.~D., \& Zhao, H. 2007, ApJ, 658, L29, \dodoi{10.1086/513509}

\bibitem[{Rix {et~al.}(2022)Rix, Chandra, Andrae, {Price-Whelan}, Weinberg, Conroy, Fouesneau, Hogg, De~Angeli, Naidu, Xiang, \& {Ruz-Mieres}}]{rixPoorOldHeart2022}
Rix, H.-W., Chandra, V., Andrae, R., {et~al.} 2022, ApJ, 941, 45, \dodoi{10.3847/1538-4357/ac9e01}

\bibitem[{Robitaille {et~al.}(2013)Robitaille, Tollerud, Greenfield, Droettboom, Bray, Aldcroft, Davis, Ginsburg, {Price-Whelan}, Kerzendorf, Conley, Crighton, Barbary, Muna, Ferguson, Grollier, Parikh, Nair, G{\"u}nther, Deil, Woillez, Conseil, Kramer, Turner, Singer, Fox, Weaver, Zabalza, Edwards, Bostroem, Burke, Casey, Crawford, Dencheva, Ely, Jenness, Labrie, Lim, Pierfederici, Pontzen, Ptak, Refsdal, Servillat, \& Streicher}]{robitailleAstropyCommunityPython2013}
Robitaille, T.~P., Tollerud, E.~J., Greenfield, P., {et~al.} 2013, A\&A, 558, A33, \dodoi{10.1051/0004-6361/201322068}

\bibitem[{{Rojas-Arriagada} {et~al.}(2014){Rojas-Arriagada}, {Recio-Blanco}, Hill, {de Laverny}, Schultheis, Babusiaux, Zoccali, Minniti, Gonzalez, Feltzing, Gilmore, Randich, Vallenari, Alfaro, Bensby, Bragaglia, Flaccomio, Lanzafame, Pancino, Smiljanic, Bergemann, Costado, Damiani, Hourihane, Jofr{\'e}, Lardo, Magrini, Maiorca, Morbidelli, Sbordone, Worley, Zaggia, \& Wyse}]{rojas-arriagadaGaiaESOSurveyMetallicity2014}
{Rojas-Arriagada}, A., {Recio-Blanco}, A., Hill, V., {et~al.} 2014, Astronomy and Astrophysics, 569, A103, \dodoi{10.1051/0004-6361/201424121}

\bibitem[{{Rojas-Arriagada} {et~al.}(2017){Rojas-Arriagada}, {Recio-Blanco}, {de Laverny}, Mikolaitis, Matteucci, Spitoni, Schultheis, Hayden, Hill, Zoccali, Minniti, Gonzalez, Gilmore, Randich, Feltzing, Alfaro, Babusiaux, Bensby, Bragaglia, Flaccomio, Koposov, Pancino, Bayo, Carraro, Casey, Costado, Damiani, Donati, Franciosini, Hourihane, Jofr{\'e}, Lardo, Lewis, Lind, Magrini, Morbidelli, Sacco, Worley, \& Zaggia}]{rojas-arriagadaGaiaESOSurveyExploring2017}
{Rojas-Arriagada}, A., {Recio-Blanco}, A., {de Laverny}, P., {et~al.} 2017, Astronomy and Astrophysics, 601, A140, \dodoi{10.1051/0004-6361/201629160}

\bibitem[{{Rojas-Arriagada} {et~al.}(2020){Rojas-Arriagada}, Zasowski, Schultheis, Zoccali, Hasselquist, Chiappini, Cohen, Cunha, {Fern{\'a}ndez-Trincado}, Fragkoudi, {Garc{\'i}a-Hern{\'a}ndez}, Geisler, Gran, Lian, Majewski, Minniti, Monachesi, Nitschelm, \& Queiroz}]{rojas-arriagadaHowManyComponents2020}
{Rojas-Arriagada}, A., Zasowski, G., Schultheis, M., {et~al.} 2020, Monthly Notices of the Royal Astronomical Society, 499, 1037, \dodoi{10.1093/mnras/staa2807}

\bibitem[{{Roman-Oliveira} {et~al.}(2023){Roman-Oliveira}, Fraternali, \& Rizzo}]{roman-oliveiraRegularRotationLow2023}
{Roman-Oliveira}, F., Fraternali, F., \& Rizzo, F. 2023, Monthly Notices of the Royal Astronomical Society, 521, 1045, \dodoi{10.1093/mnras/stad530}

\bibitem[{Saha \& Gerhard(2013)}]{sahaSecularEvolutionCylindrical2013}
Saha, K., \& Gerhard, O. 2013, Monthly Notices of the Royal Astronomical Society, 430, 2039, \dodoi{10.1093/mnras/stt029}

\bibitem[{Saha {et~al.}(2012)Saha, {Martinez-Valpuesta}, \& Gerhard}]{sahaSpinupLowmassClassical2012}
Saha, K., {Martinez-Valpuesta}, I., \& Gerhard, O. 2012, Monthly Notices of the Royal Astronomical Society, 421, 333, \dodoi{10.1111/j.1365-2966.2011.20307.x}

\bibitem[{Sch{\"o}nrich(2012)}]{schonrichGalacticRotationSolar2012a}
Sch{\"o}nrich, R. 2012, Monthly Notices of the Royal Astronomical Society, 427, 274, \dodoi{10.1111/j.1365-2966.2012.21631.x}

\bibitem[{Sch{\"o}nrich \& McMillan(2017)}]{schonrichUnderstandingInverseMetallicity2017a}
Sch{\"o}nrich, R., \& McMillan, P.~J. 2017, Monthly Notices of the Royal Astronomical Society, 467, 1154, \dodoi{10.1093/mnras/stx093}

\bibitem[{Semenov {et~al.}(2023)Semenov, Conroy, Chandra, Hernquist, \& Nelson}]{semenovFormationGalacticDisks2023}
Semenov, V.~A., Conroy, C., Chandra, V., Hernquist, L., \& Nelson, D. 2023, Formation of {{Galactic Disks I}}: {{Why}} Did the {{Milky Way}}'s {{Disk Form Unusually Early}}?, \dodoi{10.48550/arXiv.2306.09398}

\bibitem[{Shen {et~al.}(2010)Shen, Rich, Kormendy, Howard, De~Propris, \& Kunder}]{shenOURMILKYWAY2010a}
Shen, J., Rich, R.~M., Kormendy, J., {et~al.} 2010, ApJ, 720, L72, \dodoi{10.1088/2041-8205/720/1/L72}

\bibitem[{Springel {et~al.}(2021)Springel, Pakmor, Zier, \& Reinecke}]{springelSimulatingCosmicStructure2021}
Springel, V., Pakmor, R., Zier, O., \& Reinecke, M. 2021, Monthly Notices of the Royal Astronomical Society, 506, 2871, \dodoi{10.1093/mnras/stab1855}

\bibitem[{{Tepper-Garcia} {et~al.}(2021){Tepper-Garcia}, {Bland-Hawthorn}, Vasiliev, Athanassoula, Gerhard, Quillen, McMillan, Freeman, Lewis, Teyssier, Sharma, Hayden, \& Buder}]{tepper-garciaBarredMilkyWay2021}
{Tepper-Garcia}, T., {Bland-Hawthorn}, J., Vasiliev, E., {et~al.} 2021, A Barred {{Milky Way}} Surrogate from an {{N-body}} Simulation,  arXiv.
\newblock \doeprint{2111.05466}

\bibitem[{Valentini {et~al.}(2019)Valentini, Borgani, Bressan, Murante, Tornatore, \& Monaco}]{valentiniChemicalEvolutionDisc2019}
Valentini, M., Borgani, S., Bressan, A., {et~al.} 2019, Monthly Notices of the Royal Astronomical Society, 485, 1384, \dodoi{10.1093/mnras/stz492}

\bibitem[{Vasiliev(2019)}]{vasilievAGAMAActionbasedGalaxy2019}
Vasiliev, E. 2019, Monthly Notices of the Royal Astronomical Society, 482, 1525, \dodoi{10.1093/mnras/sty2672}

\bibitem[{{Vera-Ciro} {et~al.}(2014){Vera-Ciro}, D'Onghia, Navarro, \& Abadi}]{vera-ciroEffectRadialMigration2014}
{Vera-Ciro}, C., D'Onghia, E., Navarro, J., \& Abadi, M. 2014, The Astrophysical Journal, 794, 173, \dodoi{10.1088/0004-637X/794/2/173}

\bibitem[{Vickers {et~al.}(2021)Vickers, Shen, \& Li}]{vickersFlatteningMetallicityGradient2021}
Vickers, J.~J., Shen, J., \& Li, Z.-Y. 2021, ApJ, 922, 189, \dodoi{10.3847/1538-4357/ac27a9}

\bibitem[{Vislosky {et~al.}(2024)Vislosky, Minchev, Khoperskov, Martig, Buck, Hilmi, Ratcliffe, {Bland-Hawthorn}, Quillen, Steinmetz, \& {de Jong}}]{visloskyGaiaDR3Data2024}
Vislosky, E., Minchev, I., Khoperskov, S., {et~al.} 2024, Gaia {{DR3}} Data Consistent with a Short Bar Connected to a Spiral Arm,  arXiv.
\newblock \doeprint{2312.03854}

\bibitem[{Weiland {et~al.}(1994)Weiland, Arendt, Berriman, Dwek, Freudenreich, Hauser, Kelsall, Lisse, Mitra, Moseley, Odegard, Silverberg, Sodroski, Spiesman, \& Stemwedel}]{weilandCOBEDiffuseBackground1994}
Weiland, J.~L., Arendt, R.~G., Berriman, G.~B., {et~al.} 1994, The Astrophysical Journal, 425, L81, \dodoi{10.1086/187315}

\bibitem[{Weinberg {et~al.}(2017)Weinberg, Andrews, \& Freudenburg}]{weinbergEquilibriumSuddenEvents2017}
Weinberg, D.~H., Andrews, B.~H., \& Freudenburg, J. 2017, ApJ, 837, 183, \dodoi{10.3847/1538-4357/837/2/183}

\bibitem[{White \& Springel(2000)}]{10.1007/10719504_62}
White, S. D.~M., \& Springel, V. 2000, in The First Stars, ed. A.~Weiss, T.~G. Abel, \& V.~Hill (Berlin, Heidelberg: Springer Berlin Heidelberg), 327--335

\bibitem[{Wylie {et~al.}(2021)Wylie, Gerhard, Ness, Clarke, Freeman, \& {Bland-Hawthorn}}]{wylieA2A210002021}
Wylie, S.~M., Gerhard, O.~E., Ness, M.~K., {et~al.} 2021, arXiv e-prints, arXiv:2106.14298

\bibitem[{Xiang \& Rix(2022)}]{xiangTimeresolvedPictureOur2022}
Xiang, M., \& Rix, H.-W. 2022, Nature, 603, 599, \dodoi{10.1038/s41586-022-04496-5}

\bibitem[{Zoccali {et~al.}(2008)Zoccali, Hill, Lecureur, Barbuy, Renzini, Minniti, G{\'o}mez, \& Ortolani}]{zoccaliMetalContentBulge2008a}
Zoccali, M., Hill, V., Lecureur, A., {et~al.} 2008, Astronomy and Astrophysics, 486, 177, \dodoi{10.1051/0004-6361:200809394}

\bibitem[{Zoccali {et~al.}(2014)Zoccali, Gonzalez, Vasquez, Hill, Rejkuba, Valenti, Renzini, {Rojas-Arriagada}, {Martinez-Valpuesta}, Babusiaux, Brown, Minniti, \& McWilliam}]{zoccaliGIRAFFEInnerBulge2014}
Zoccali, M., Gonzalez, O.~A., Vasquez, S., {et~al.} 2014, A\&A, 562, A66, \dodoi{10.1051/0004-6361/201323120}

\bibitem[{Zoccali {et~al.}(2017)Zoccali, Vasquez, Gonzalez, Valenti, {Rojas-Arriagada}, Minniti, Rejkuba, Minniti, McWilliam, Babusiaux, Hill, \& Renzini}]{zoccaliGIRAFFEInnerBulge2017}
Zoccali, M., Vasquez, S., Gonzalez, O.~A., {et~al.} 2017, Astronomy and Astrophysics, 599, A12, \dodoi{10.1051/0004-6361/201629805}

\end{thebibliography}
\bibliographystyle{aasjournal}



\end{document}